\documentclass{aa}

\usepackage{graphicx}
\usepackage{txfonts}
\usepackage{lipsum}
\usepackage{multirow}
\usepackage{subcaption}         
\usepackage{lscape}             
\usepackage{placeins}           
\usepackage[colorlinks=true, linkcolor=blue, citecolor=blue, urlcolor=blue]{hyperref}

\begin{document}

   \title{Influence of Barlens on the Bulge Parameters in the 2D Image Decomposition}

   \author{Xinyang Li\inst{1,2}
        \and Zhao-Yu Li\inst{1,2}\fnmsep\thanks{lizy.astro@sjtu.edu.cn}
        \and Yang A. Li\inst{1,2}\fnmsep\thanks{yangli\_anpin@sjtu.edu.cn}
        \and Ming-Yang Zhuang\inst{3}
        \and Xiaojie Liao\inst{1,2}
        }

   \institute{Department of Astronomy, School of Physics and Astronomy, Shanghai Jiao Tong University, 800 Dongchuan Road, Shanghai 200240, China
            \and State Key Laboratory of Dark Matter Physics, School of Physics and Astronomy, Shanghai Jiao Tong University, Shanghai 200240, China
            \and Department of Astronomy, University of Illinois Urbana-Champaign, Urbana, 61801, IL, USA\\}

   \date{Accepted September 30, 20XX}

 
  \abstract
   {Recent observations and simulations have shown that a buckled bar in the face-on view can be considered as a combination of a long flat bar and a short round barlens (corresponding to the boxy/peanut bulge in the edge-on view). However, the barlens component can be misidentified as the bulge, potentially leading to inaccurate bulge parameter measurements in two-dimensional (2D) image decomposition.}
   {Our goal is to explore the optimal method for modeling the barlens component and to understand its impact on bulge parameter measurements in 2D image decomposition. } 
   {We first analyze mock images from two different simulations (with/without bulge) to verify our decomposition method. We then apply the method to two nearby barred galaxies, NGC 1533 and NGC 7329, from the Carnegie-Irvine Galaxy Survey (CGS). Using GALFIT, we conduct 2D image decomposition by gradually increasing the complexity of model configurations. We also explore the effects of inclination by projecting the simulated galaxy to various viewing angles and analyzing the variations in bulge and barlens parameters. }
   {From the mock images, we find that the bulge-to-total ratio (B/T) could be overestimated by 50$\%$ without considering the barlens component; the Sérsic index and effective radius of the bulge are also affected to varying degrees. The decomposition results of the two CGS galaxies are consistent with our mock image tests. Uncertainties of the structural parameters of the bulge and barlens are larger at higher inclination angels due to the strong projection effect in the central region.}
   {Our findings underscore the necessity of accurately modeling the barlens, revealing that its inclusion in 2D image decomposition can lead to a decrease in B/T by $\sim$30-50$\%$, with other bulge parameters, such as the Sérsic index and effective radius, also affected. }


   \maketitle

\section{Introduction} \label{sec:intro}

The bulge in the central region of a disk galaxy is a crucial component for understanding the formation and evolution of disk galaxies. Recent studies have shown the diverse nature of bulges, which can be classified into classical bulges and pseudobulges. Classical bulges, often described as mini-ellipticals at the centers of disks, are believed to form early through violent merger processes, offering vital insights into the early stages of galaxy formation \citep{Eggen1962, Toomre1972, Gott1977, Bender1992, Renzini1999}. In contrast, pseudobulges exhibit disky shapes, relatively young stellar populations, flat luminosity distributions, and rotationally dominated kinematics \citep{Kormendy2004, Gadotti2009, Kormendy2013}. These features suggest that pseudobulges form over long timescales via secular processes, where non-axisymmetric structures such as bars and spiral arms drive gas inflow toward the galaxy center, fueling central star formation \citep{Sellwood1993, Kormendy2004, Sellwood2014, Tonini2016}. Distinguishing between classical and pseudobulges is crucial for a deeper understanding of disk galaxy evolution.

In the local Universe, a significant fraction of disk galaxies ($\sim$60$\%$) host bars in their central regions  \citep{Eskridge2000, Mendez2007, Erwin2018}, which can influence the measurement of bulge parameters in 2D image decomposition \citep{Laurikainen2004, Gadotti2008, Fisher2008}. Both observations \citep{Lutticke2000, Laurikainen2014, Erwin2017, Blana2017, Gadotti2020} and simulations \citep{Raha1991, Berentzen1998, Martinez2004, Li2015, Fragkoudi2017} suggested that the bar consists of two components: an elongated thin bar and a thickened inner structure. In edge-on views, this inner component appears boxy, peanut-shaped, or X-shaped, and is referred to as the boxy/peanut (B/P) bulge, a particular type of pseudobulge. In face-on views, the B/P bulge appears as a lens-like oval structure known as the barlens, which typically extends to about half the length of the bar and has lower ellipticity \citep{Laurikainen2011, Athanassoula2015, Buta2015, Laurikainen2017}. Recent studies have shown that the fraction of B/P bulges in barred galaxies in the local Universe ranges from 50–70\% \citep{Erwin2017, Kruk2019, Gadotti2020}. \citet{Li2017} and \citet{Erwin2017} further demonstrated that the presence of a B/P bulge is strongly correlated with galaxy stellar mass: the majority ($\sim80\%$) of barred galaxies with log($M_\star/M_\odot$) $\geq$ 10.4 host B/P bulges, compared to only $\sim20\%$ B/P bulge fraction among lower-mass galaxies. The formation mechanisms of the B/P bulge could be the buckling instability of the bar \citep{Raha1991, Merritt1994, Martinez2004, Smirnov2019}, or the vertical resonances of the stellar orbits within the bar \citep{Combes1981, Combes1990, Quillen2014, Sellwood2020}. 

The barlens, which resides in the central region of a galaxy, can easily be mistaken as part of the bulge in face-on views or at moderate inclinations. Previous studies typically consider three components in disk galaxy image decomposition: the bulge, bar, and disk. Ignoring the barlens component would result in attributing its flux to the bulge, leading to an overestimation of the bulge-to-total (B/T) ratio and Sérsic index ($n$) \citep{Laurikainen2018, Erwin2021}. \citet{Athanassoula2015} performed 2D image decomposition on $N$-body simulations of barred disks to derive the parameters of the barlens. Using generalized elliptical isophotes for all components, they found that the Sérsic index of the barlens is less than 2, and it accounts for 0.13–0.33 of the galaxy's total flux. The barlens ellipticity ranges from 0.7 to 1.0, and its length is 0.4–0.8 times the length of the bar, with only a weak dependence on inclination angle. In a 2D image decomposition of galaxies from the Calar Alto Legacy Integral Field Area (CALIFA) survey, \citet{Laurikainen2018} showed that the barlens contributes approximately 13\% of the total galaxy light, with the remaining flux primarily attributed to a possible distinct bulge component that contributes $\leq$10\%. They also found that the barlens typically exhibits an exponential profile ($n \leq 1$).

In previous work, the influence of modeling different galaxy structures, including the bar, ring, spiral arms and lens, on the bulge parameters have been extensively studied \citep{Gadotti2008, Gao2017, Chugunov2023, Marchuk2024}. \citet{Athanassoula2015} have proposed methods to measure the barlens component in 2D image decomposition. However, without gradually increasing the modeling complexity, their modeling strategy could not properly answer this important question: what is the impact of modeling the barlens on the bulge parameters? Built upon the previous efforts in the barlens modeling, here we aim to develop an optimal strategy for modeling the barlens in 2D image decomposition and to quantify its influence on the bulge measurements. We carefully design a series of decomposition tests that minimize parameter degeneracies between the overlapping structures (i.e., bar, bulge and barlens). To better constrain the intrinsic parameters of the barlens, we first use mock images from two different simulations: a pure-disk $N$-body simulation featuring a buckled bar without a pre-existing classical bulge, and a Milky Way-like galaxy from the Auriga simulation with the bulge built in the previous merger process. Afterwards, we apply the same 2D image decomposition strategy to real observational images of two nearby barred galaxies that exhibite the barlens structure. Unlike previous works (e.g., \citealt{Athanassoula2015}), which primarily emphasized the identification and characterization of the barlens component itself, our primary focus here is to study the influence of the barlens modeling on the bulge parameters. We gradually increase the complexity of the model to better understand the variation of the bulge parameters after the barlens component is included. Moreover, we also investigate the effect of the disk inclination on the structural decomposition results, a factor that has not been thoroughly explored in previous barlens studies.

This paper is organized as follows: In Section~\ref{sec:data}, we provide an overview of the mock images, observational data, and the decomposition method used to model the barlens component. The results of the mock image decomposition are presented in Section~\ref{sec:result}, along with applications to real observations. The effects of inclination angle and other statistical tests on the decomposition results are discussed in Section~\ref{sec:disc}. The summary of our findings is given in Section~\ref{sec:conc}.

\section{Data and Method} \label{sec:data}

\subsection{Mock Image Creation from Simulation}

To first understand the intrinsic properties of the barlens, we begin with a mock galaxy image generated from a pure-disk galaxy simulation without any pre-existing classical bulge (hereafter Model 1). The initial conditions (ICs) are set up following \citet{Tepper2021}, with a dark matter halo of mass $M_{\rm h}\sim 1.2\times10^{12}M_\odot$, a stellar disk of mass $M_{\rm d} = 4.3\times10^{10}M_\odot$, a scale length of 2.5 kpc, and a scale height of 0.3 kpc (resembling a thin disk). Both the dark matter halo and the disk are modeled with $10^6$ particles. The ICs are generated using AGAMA \citep{Vasiliev2019} and evolved with GADGET4 \citep{Springel2021} for 5 Gyr. During the evolution of Model 1, a bar forms at $\sim$1.1 Gyr and buckles into a peanut-shaped bulge at $\sim$2.3 Gyr.

Next, to evaluate the impact of the barlens on the bulge measurement, we use a simulated galaxy from the Auriga project\footnote{\url{https://wwwmpa.mpa-garching.mpg.de/auriga/data}} \citep{Grand2017, Grand2024}. The Auriga project includes 30 Milky Way-like galaxies simulated within the framework of $\Lambda$CDM cosmology using the moving mesh code AREPO \citep{Springel2010}. The simulations account for various physical processes, such as star formation, chemical enrichment, and feedback effects. Further details can be found in \citet{Grand2017}. The Auriga project successfully forms a series of barred galaxies. After visually checking the Auriga simulations, we decide to use the galaxy Auriga 6 at $z = 0$ (hereafter Model 2) to generate mock images, which has a clear bar with a prominent B/P bulge from the edge-on view. The galaxy also features a ring structure surrounding the bar and weak spiral arms in its outer disk. 

\begin{figure}[t]
\includegraphics[width=\columnwidth]{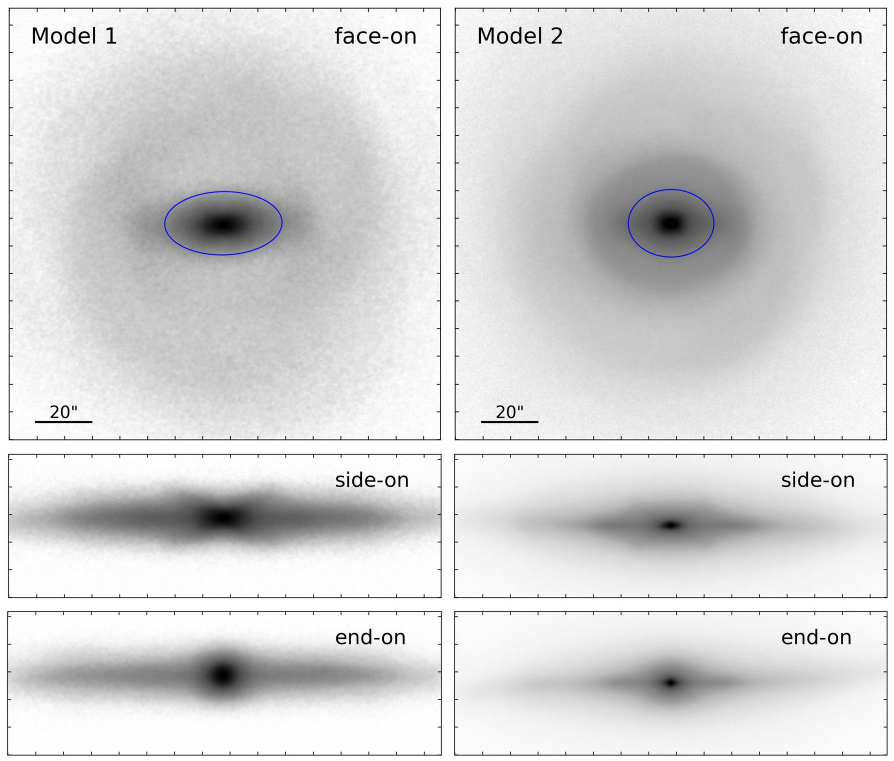}
    \caption{Mock images of Model 1 and Model 2 in the face-on view (top), side-on view (middle), and end-on view (bottom). The barlens and the corresponding B/P bulge can be seen clearly in the face-on view and edge-on views, respectively. The blue ellipses in the top panels show the visually identified outer edge of the barlenses. }
    \label{fig:Au6}
\end{figure}

Similar to previous works \citep{Erwin2013, Athanassoula2015, Anderson2022}, we create mock images of the simulations based on the stellar mass distribution. The particles are projected onto a 2D grid of pixels, with each pixel value representing the sum of the masses of all the stellar particles within that pixel. To simulate observations from ground-based telescopes such as the Carnegie-Irvine Galaxy Survey \citep[CGS,][]{Ho2011, Li2011}, we assume that the galaxy is located at a distance of 40 Mpc. The simulations are projected in the face-on view (in the $X$-$Y$ plane), with a pixel scale of approximately 0.26 arcsec pixel$^{-1}$, or 0.048 kpc pixel$^{-1}$. The image is then convolved with the point spread function (PSF) from the $R$-band CGS images\footnote{The $R$-band images are used considering the relative deep image, high spatial resolution, and less sensitivity to the dust absorption and the star forming regions \citep{Huang2013}.}. The total flux of the convolved image is then scaled to match the total flux of the $R$-band images of nearby CGS disk galaxies, with the total apparent magnitude of $\sim$ 11.6 mag. Poisson noise and sky background are then added to the image, with the sky background modeled as a Gaussian distribution, where the mean and standard deviation of the sky pixels are determined based on the sky values of the $R$-band CGS images.

The mock images of the two simulations in face-on, side-on, and end-on views are shown in Figure~\ref{fig:Au6}. In the side-on view, the B/P bulge is visible, corresponding to the barlens in the face-on view. To fully explore how the disk inclination angle affects the measurements of the barlens and the bulge parameters, we then generate mock images for different disk inclinations ($i = 30^\circ$, 45$^\circ$, 60$^\circ$) and various position angles of the bar (PA$_{\rm bar}$ = 0$^\circ$, 30$^\circ$, 60$^\circ$, and 90$^\circ$).

\begin{figure}[t]
\includegraphics[width=\columnwidth]{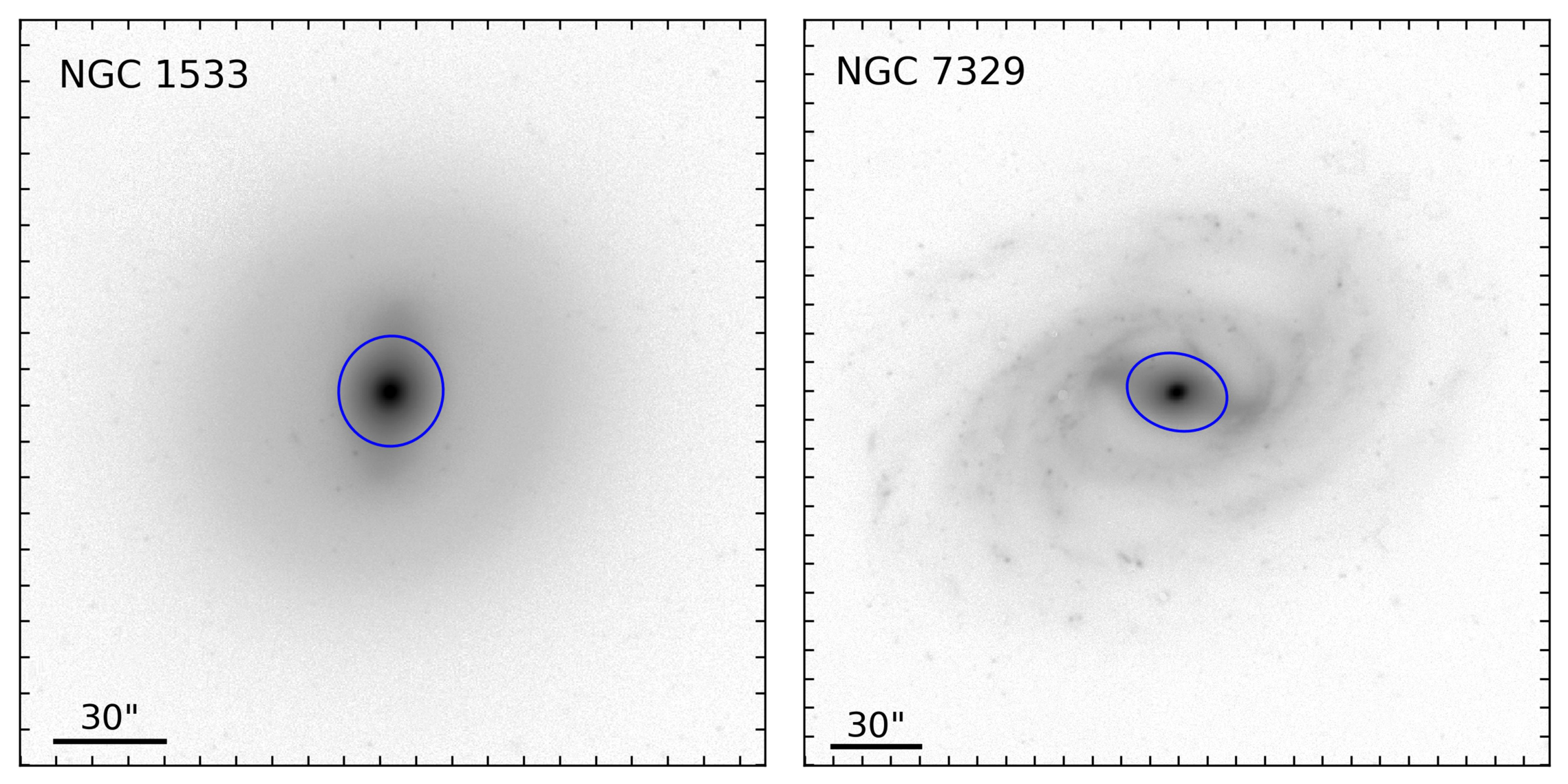}
    \caption{$R$-band star-cleaned images of NGC 1533 (left) and NGC 7329 (right) with the barlens component identified in \citet{Li2017}. The blue ellipses show the visually identified outer edge of the barlenses. }
    \label{fig:ngc}
\end{figure}

\subsection{Observational Images from the CGS Survey}

After testing the barlens decomposition method with mock images, we apply the same method to two nearby barred galaxies (NGC 1533 and NGC 7329) in the CGS survey that exhibit clear barlens structures \citep{Li2017}. The $R$-band images of these galaxies are shown in Figure~\ref{fig:ngc}. The images have a pixel scale of 0.26 arcsec pixel$^{-1}$, with a median seeing of 1.01 arcsec and a surface brightness depth of 26.4 mag arcsec$^{-2}$. In our analysis, we use star-cleaned images and PSF images from \citet{Li2011}. The star-cleaned images are free from contamination by foreground stars and background galaxies. In the following 2D image decomposition, an additional mask image is not needed. Comprehensive 2D image decompositions, including the bulge, bar, and ring components, have been carried out for these galaxies by \citet{Gao2017}. However, their work did not take into account the barlens as an additional component, which could potentially influence the bulge properties.

\subsection{Image Decomposition Strategy} 

\subsubsection{Light profile functions and isophotal shapes}

We use GALFIT to perform the 2D image decomposition \citep{Peng2010}. GALFIT provides a variety of analytic functions for modeling light profiles, including Sérsic \citep{Sersic1963}, exponential, modified Ferrer, Moffat, and King functions. Additionally, the isophotal shape can be modeled using Fourier modes to capture features such as the B/P-shaped isophote, enabling more realistic galaxy models. The weight map used in the fitting was automatically generated by GALFIT.

We adopt the exponential profile to model the disk component. Sérsic profiles are used to model the bulge, bar, and barlens. Extensive experimentation reveals that using a Sérsic profile instead of a modified Ferrer function to model the bar does not significantly influence the bulge-to-total ratio (B/T), Sérsic index ($n$), and effective radius ($r_{\rm e}$) of the bulge \citep{Gao2017}. For simplicity, we use the Sérsic profile in this work.

Following \citet{Gao2017}, we use two broken exponential profiles to model the ring or ring-like structures, with one component truncated in the inner region and the other in the outer region. The overlap between these two components creates a bump in the surface brightness profile, corresponding to the ring. For both the mock images and CGS observations, the truncation parameters for the disk component are also set following \citet{Gao2017}. We also tested modeling the ring using an exponential disk combined with a Gaussian ring profile \citep{Erwin2015}; the results, presented in Appendix~\ref{apx2}, are consistent with those obtained using two broken exponential profiles.

The barlens component becomes B/P-shaped at relatively large inclinations ($i \geq 45^\circ$) and for a side-on bar (PA$_{\rm bar}$ $\geq 60^\circ$). To model the barlens in this case, we use the Fourier mode in GALFIT to generate a boxy/X-shaped isophote \citep{Peng2010}, adopting the $m=4$ mode.

\subsubsection{General procedure of the 2D image decomposition} \label{subsubsec:strategy}

When performing the 2D image decomposition, the initial parameters of each component are important. Following the fitting strategy in \citet{Gao2017}, we start from the simplest fitting configuration with the bar/disk components (bar+D) for Model 1 and the bulge/disk components (B+D) for Model 2. Then we gradually increase the complexity by adding additional components to the image decomposition. The best-fit parameters of the preceding configurations are used as the initial input parameters for the same component of the subsequent configuration. For Model 1, the second fitting configuration, denoted as bar/disk1/disk2 (bar+D1+D2), employs a pair of broken exponential profiles for the disk to account for the ring. Then the barlens component is added in the third fitting configuration denoted as barlens/bar/disk1/disk2 (bl+bar+D1+D2). In addition, we also consider a fourth fitting configuration, with a single exponential disk rather than the two broken exponential disks, namely, barlens/bar/disk (bl+bar+D). For Model 2, the second fitting configuration includes a bar component in addition to the B+D, i.e., the bulge/bar/disk configuration (B+bar+D). In the third fitting configuration, we again use a pair of broken exponential profiles to model the ring structure at the bar end, and denote it as bulge/bar/disk1/disk2 (B+bar+D1+D2). In the last fitting configuration, we add the barlens component, denoted as bulge/bar/barlens/disk1/disk2 (B+bar+bl+D1+D2). 

For both Model 1 and Model 2, the parameters of the components in different fitting configurations are all set free, except for the barlens, which requires special handling. Following \citet{Athanassoula2015}, we visually identify the outer boundary of the barlens from the galaxy morphology that appears as a rounder structure extending to about half of the elongated bar. Moreover, to further confirm the position of the outer edge, we also inspect the 1D surface brightness profile along the minor axis of the bar, where a noticeable change in the light profile slope typically marks the transition between the barlens and the surrounding disk. We then fit an ellipse to the visually identified edge of the barlens, and denote its semi-major axis as $a_{\rm bl}$. According to \citet{Laurikainen2018}, the effective radius of the barlens is then fixed to approximately 0.4$a_{\rm bl}$. In Model 2, the bulge overlaps with both the barlens and the bar, leading to significant parameter degeneracy. To address this, additional constraints are applied to the barlens component. Initially, all parameters are set free in the B+bar+bl+D1+D2 configuration fro Model 2 to obtain the initial guess values. The effective radius and axial ratio of the barlens are then constrained based on the visually identified outer edge of the barlens ellipse. Moreover, the Sérsic index of the barlens is fixed to the value obtained from Model 1 ($n = 1.0$), with alternative values ($n = 0.5$ and 1.5) tested to assess their impact on the bulge parameters. The final Sérsic index for Model 2 is the one that can result in the smallest residual map. The decomposition process proceeds iteratively: first, the parameters of the bulge, bar, and disk are fixed in turn to break the degeneracy and find reasonable values for each component. Once this is done, all parameters of the disk, bar, and bulge are freed in the B+bar+bl+D1+D2 configuration and the fitting is performed again to obtain the final best-fit parameters for each component.\footnote{In some cases, certain parameters, such as the broken radius and ellipticity of the broken exponential disk, may still need to be fixed to avoid unreasonable results.}

In contrast to \citet{Athanassoula2015}, in our fitting procedure, the size, axial ratio, and Sérsic index of the barlens are always kept fixed. If the size of the barlens is not constrained, the best-fit value may become unreasonable, e.g., the barlens covers nearly the entire length of the bar. Similarly, if the Sérsic index of the barlens is not fixed, it would likely become excessively high due to strong degeneracy with both the bulge and the bar.

\section{Results} \label{sec:result}

\subsection{Mock Image Decomposition of Model 1} \label{subsec:m1}

\begin{figure*}[ht!]
\includegraphics[width=\textwidth]{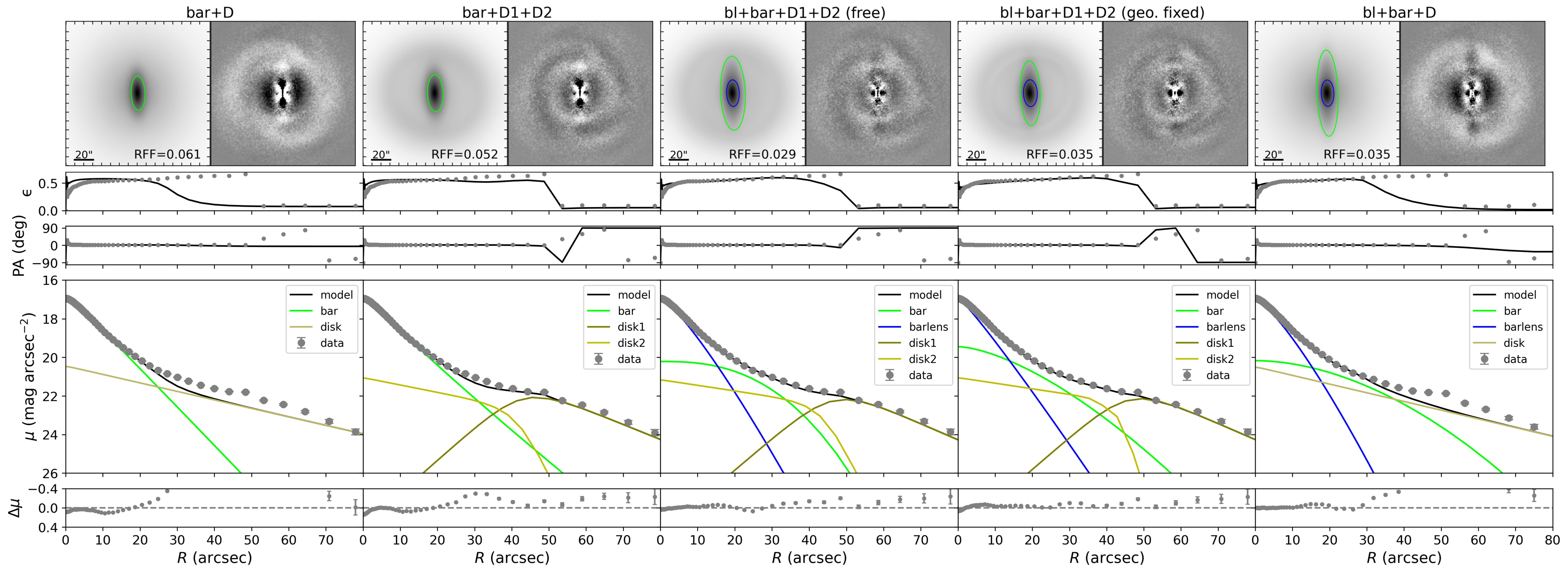}
    \caption{2D image decomposition results of the Model 1 mock image in the face-on view. In each column, the top row shows the best-fit 2D model image (left) and the residual image (right). The green and blue ellipses indicate the outlines of the bar and barlens, respectively, with semi-major axis of 2$r_{\rm e}$. The other rows show results of the isophotal analysis, including the radial profiles of ellipticity (second row), the position angle (third row), the surface brightness (forth row), and the residual between data and the best-fit model (bottom row). From left to right, the fitting configuration are bar+D, bar+D1+D2, bl+bar+D1+D2 (free parameters fitting), and bl+bar+D1+D2 (fixing the size and axial ratio of the barlens), respectively (see details in the text). The fifth column (bl+bar+D) shows the result of the fitting with only one exponential disk. The RFF of each configuration is labeled in the model image. }
    \label{fig:nbody_decomposition}
\end{figure*}

\begin{table*}[ht!]
	\caption{Best-fit parameters of different components from 2D image decomposition of the Model 1 mock image. The columns from left to right show the component, the decomposition configuration, the Sérsic index, the effective radius, the surface brightness at the effective radius, the axial ratio, the position angle and the relative flux ratio of the component, respectively. }
\label{tab:nbody_2D_params}
\resizebox{517pt}{!}{
	\begin{tabular}{cccccccc}
		\hline
		Component & Configuration & $n$ & $r_{\rm e}$ & $\mu_{\rm e}$ & b/a & PA & $f/f_{\rm tot}$ \\
             &  &  & (arcsec) & (mag arcsec$^{-2}$) &  & ($^\circ$) &  \\
             (1) & (2) & (3) & (4) & (5) & (6) & (7) & (8) \\ 
		\hline
		\multirow{3}{*}{barlens} & \multicolumn{1}{l}{bl+bar+D1+D2 (free)} & $0.83\pm 0.02$ & $7.22\pm 0.14$ & $18.28\pm 0.02$ & $0.48\pm 0.01$ & $2.28\pm 0.12$ & $0.33\pm 0.02$ \\
        & \multicolumn{1}{l}{bl+bar+D1+D2 (geo. fixed)} & $0.93\pm 0.02$ & $7.25\pm 0.50$ (fixed) & $18.48\pm 0.01$ & $0.52\pm 0.02$ (fixed) & $2.88\pm 0.06$ & $0.32\pm 0.02$ \\
        & \multicolumn{1}{l}{bl+bar+D} & $0.79\pm 0.04$ & $7.14\pm 0.23$ & $18.29\pm 0.03$ & $0.48\pm 0.01$ & $2.64\pm 0.02$ & $0.34\pm 0.03$ \\
        \hline
		\multirow{5}{*}{bar} & \multicolumn{1}{l}{bar+D} & $0.98 \pm 0.07$ & $9.29\pm 0.57$ & $18.47\pm 0.07$ & $0.41\pm 0.01$ & $1.97\pm 0.04$ & $0.45\pm 0.06$ \\
        & \multicolumn{1}{l}{bar+D1+D2} & $1.10\pm 0.02$ & $9.96\pm 0.08$ & $18.60 \pm 0.02$ & $0.43\pm 0.01$ & $1.89\pm 0.02$ & $0.48\pm 0.04$ \\
        & \multicolumn{1}{l}{bl+bar+D1+D2 (free)} & $0.38\pm 0.04$ & $20.04\pm 0.64$ & $20.72\pm 0.07$ & $0.34\pm 0.01$ & $1.28\pm 0.12$ & $0.14\pm 0.02$ \\
        & \multicolumn{1}{l}{bl+bar+D1+D2 (geo. fixed)} & $0.65\pm 0.06$ & $17.56\pm 0.48$ & $20.50\pm 0.04$ & $0.31\pm 0.05$ & $0.94\pm 0.19$ & $0.15\pm 0.02$ \\
        & \multicolumn{1}{l}{bl+bar+D} & $0.54\pm 0.06$ & $23.32\pm 0.61$ & $20.99\pm 0.06$ & $0.24\pm 0.01$ & $0.66\pm 0.12$ & $0.13\pm 0.02$\\
        \hline
	\end{tabular}
}
\tablefoot{In column (2), `free' stands for freeing all parameters of the barlens; `geo. fixed' stands for fixing the axial ratio and the size of the barlens. }
\end{table*}

Model 1 is a pure barred disk galaxy without a pre-existing bulge component. Therefore, during the modeling process, we only consider the bar, barlens, and disk components. Figure~\ref{fig:nbody_decomposition} shows the results of the 2D image decomposition for the face-on mock image of Model 1. The best-fit parameters for the barlens and bar components across different fitting configurations are listed in Table~\ref{tab:nbody_2D_params}. Following \citet{Gao2017}, the errors for the free parameters of the components are calculated using the Monte Carlo method. This involves perturbing the sky background 100 times, with the perturbation drawn from a normal distribution with a standard deviation of 25.13 mag arcsec$^{-2}$ (twice the standard deviation of the sky background). The error for each parameter is then determined as the standard deviation of the best-fit parameters from these perturbed images, excluding any failed or unreasonable fitting results. For the fixed parameters (e.g., the barlens axial ratio and effective radius $r_{\rm e}$), we measure the geometric parameters of the shape of the barlens 50 times and use the standard deviation of the parameters as the corresponding uncertainties.

To quantify the improvement in the residuals within the barlens region, we use the residual flux fraction (RFF) parameter \citep{Hoyos2011} defined as: 
\begin{equation}
    {\rm RFF}=\frac{\sum_{i,j\in \rm A}|I_{i,j}-I_{i,j}^{\rm model}|-0.8\times \sum_{i,j\in \rm A}\sigma_{i,j}}{\sum_{i,j\in \rm A}I_{i,j}}, 
	\label{eq:rff}
\end{equation}
where the $A$ parameter represents the area of the image. Since we are interested in the residual improvement in the central barlens region, we choose the visually identified area of the barlens for the $A$ parameter (for Model 1, see top left panel of Figure~\ref{fig:Au6}). $I_{i,j}$ is the pixel value of the mock image and $|I_{i,j}-I_{i,j}^{\rm model}|$ is the absolute value of the residual at each pixel. $\sigma_{i,j}$ is the background standard dispersion. Compared to traditional $\chi^2$ method, RFF offers a noise-corrected measure of residuals; a perfect model with residuals arising solely from background Gaussian noise should yield ${\rm RFF} \sim 0$. 

To fit the mock image, we begin with the bar+D configuration, with the results shown in the first column of Figure~\ref{fig:nbody_decomposition}. The residual image for the bar+D configuration shows relatively large residuals in both the central bar region and the outer disk. This is further highlighted in the bottom panel, which shows the surface brightness profile residuals. In the second column (bar+D1+D2), we use a pair of broken exponential disks to model the ring-like structure outside the bar. This significantly improves the residuals in the disk region ($R > 25$ arcsec). However, there is little change in either the bar parameters or the residual map in the central bar region ($R < 20$ arcsec). The RFF parameter decreases slightly from 0.061 to 0.052. The Sérsic index of the bar ($n_{\rm bar}$) is approximately 1, indicating a near-exponential surface brightness profile in the galaxy's central region. From the residual map, it is evident that both ends of the bar are not properly modeled, suggesting that the best-fit bar configuration does not adequately represent the overall bar light profile.

After adding the barlens component, i.e., in the bl+bar+D1+D2 configuration (Figure~\ref{fig:nbody_decomposition}, third column), there is a significant improvement in both the central region of the residual map and the residual surface brightness profile. RFF decreases from approximately $\sim$0.05 to $\sim$0.03. Additionally, $n_{\rm bar}$ is reduced from $\sim$1 to $\sim$0.5, resulting in a flatter profile. The effective radius of the bar ($r_{\rm e, bar}$) increases from around $\sim$10 arcsec to $\sim$20 arcsec. The Sérsic index of the barlens ($n_{\rm bl}$) is $\sim$0.8, which is close to an exponential profile and steeper than the bar. The barlens is shorter than the bar, with $r_{\rm e, bl} \sim 0.4 r_{\rm e, bar}$. These results for the Sérsic index and relative length of the barlens are consistent with previous studies \citep{Athanassoula2015, Laurikainen2018}, confirming the reliability of our decomposition method. This also highlights the necessity of modeling the buckled bar with a short barlens (exponential profile) and a long bar (flattened profile). For the disk, the choice between a single exponential disk and a pair of broken exponential disks has little impact on the barlens parameters. After including a barlens component, the scale length of the inner disk increases, and its central surface brightness becomes fainter, while the parameters of the outer disk remain roughly unchanged. Details of the disk parameters can be found in Appendix~\ref{apx3}. 

In another test, we fix the size and the axial ratio of the barlens in the bl+bar+D1+D2 configuration (Figure~\ref{fig:nbody_decomposition}, fourth column), using parameters obtained by fitting an ellipse to the visually identified outer edge of the barlens (blue ellipse in the top left panel of Figure~\ref{fig:Au6}). Specifically, we fix the effective radius of the barlens ($r_{\rm e,bl}$) to 7.25 arcsec ($\sim 0.4a_{\rm bl}$, according to \citealt{Laurikainen2018})\footnote{$a_{\rm bl}$ is the semi-major axis of the best-fit ellipse} and set the axial ratio to 0.52. Overall, the best-fit parameters show only small changes, with a slight increase in both $n_{\rm bl}$ and $n_{\rm bar}$. Moreover, RFF remains virtually unaffected by fixing the geometric parameters of the barlens. This test confirms that our fitting strategy, which fixes the barlens parameters to break degeneracies, is reasonable. To further evaluate whether the barlens parameters are influenced by the method used to model the disk, and to ensure that the improvement in the central residual primarily results from adding the barlens component rather than from the more complex disk modeling, we also perform a fit with the bl+bar+D configuration (the fifth column of Figure~\ref{fig:nbody_decomposition}). The result shows that the choice between an exponential disk and a pair of broken exponential disks has little effect on the barlens parameters. Additionally, RFF remains essentially unchanged compared to the bl+bar+D1+D2 configuration in the fifth column, confirming that the improvement in the central residual is primarily due to the inclusion of the barlens component.

\subsection{Mock Image Decomposition of Model 2} \label{subsec:m2}

\begin{figure*}[ht!]
\includegraphics[width=\textwidth]{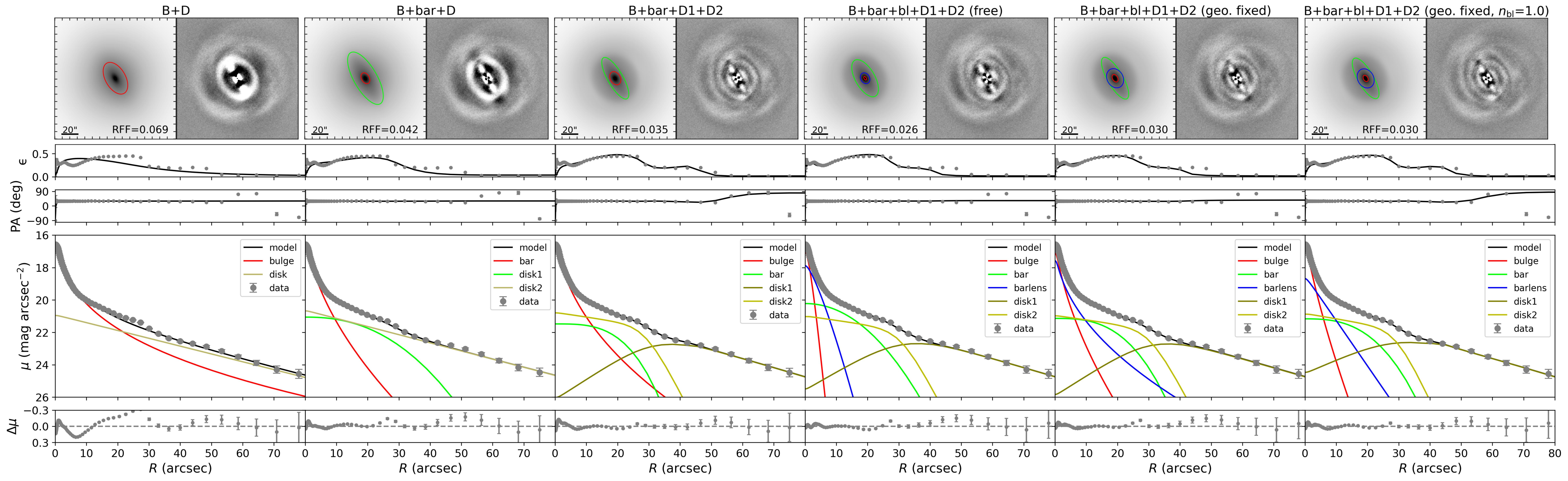}
    \caption{2D image decomposition results of the Model 2 mock image in the face-on view. The red, green, and blue ellipses indicate the outlines of the bulge, bar, and barlens, respectively, with semi-major axis of 2$r_{\rm e}$. Other conventions are similar to Figure~\ref{fig:nbody_decomposition}. The components used in the fitting configurations are shown on the top. }
    \label{fig:au6_faceon_decomposition}
\end{figure*}

\begin{table*}[ht!]
	\caption{Best-fit parameters of different components from 2D image decomposition of the Model 2 mock image. The meaning of each column is shown in Table~\ref{tab:nbody_2D_params}. }
\label{tab:au6_faceon_2D_params}
\resizebox{517pt}{!}{
	\begin{tabular}{cccccccc}
		\hline
		Component & Configuration & $n$ & $r_{\rm e}$ & $\mu_{\rm e}$ & b/a & PA & $f/f_{\rm tot}$ \\
             &  &  & (arcsec) & (mag arcsec$^{-2}$) &  & ($^\circ$) &  \\
             (1) & (2) & (3) & (4) & (5) & (6) & (7) & (8) \\ 
		\hline
		\multirow{6}{*}{bulge} & \multicolumn{1}{l}{B+D} & $3.41 \pm 1.06$  & $11.21\pm 5.47$ & $20.46\pm 1.13$ & $0.58\pm 0.03$ & $29.79\pm 0.49$ & $0.45\pm 0.14$ \\
        & \multicolumn{1}{l}{B+bar+D} & $1.81 \pm 0.32$  & $3.77 \pm 0.71$ & $18.79 \pm 0.32$ & $0.67\pm 0.02$ & $30.67 \pm 0.16$ & $0.21\pm 0.03$ \\
        & \multicolumn{1}{l}{B+bar+D1+D2} & $2.15 \pm 0.32$ & $4.55\pm 1.10$ & $19.15 \pm 0.32$ & $0.71\pm 0.01$ & $30.38 \pm 0.20$ & $0.25 \pm 0.02$ \\
        & \multicolumn{1}{l}{B+bar+bl+D1+D2 (free)} & $0.83 \pm 0.05$ & $1.29 \pm 0.04$ & $17.60 \pm 0.12$ & $0.53 \pm 0.02$ & $29.52 \pm 0.07$ & $0.05 \pm 0.01$ \\
        & \multicolumn{1}{l}{B+bar+bl+D1+D2 (geo. fixed)} & $1.71 \pm 0.13$ & $2.65\pm 0.24$ & $18.93\pm 0.35$ & $0.63 \pm 0.03$ & $30.01\pm 0.62$ & $0.08\pm 0.02$ \\
        & \multicolumn{1}{l}{B+bar+bl+D1+D2 (geo. fixed, $n_{\rm bl}$=1.0)} & $1.34 \pm 0.04$ & $2.10 \pm 0.06$ & $18.19 \pm 0.05$ & $0.65 \pm 0.01$ & $29.06 \pm 0.08$ & $0.10 \pm 0.01$ \\
        \hline
		\multirow{5}{*}{bar} & \multicolumn{1}{l}{B+bar+D} & $0.43 \pm 0.04$ & $19.03 \pm 3.52$ & $21.66\pm 0.55$ & $0.39 \pm 0.15$ & $28.86 \pm 0.74$ & $0.12\pm 0.03$ \\
        & \multicolumn{1}{l}{B+bar+D1+D2} & $0.30\pm 0.05$ & $15.25\pm 0.55$ & $21.83\pm 0.13$ & $0.32\pm 0.02$ & $30.26\pm 0.15$ & $0.05\pm 0.02$ \\
        & \multicolumn{1}{l}{B+bar+bl+D1+D2 (free)} & $0.48 \pm 0.09$ & $13.35 \pm 0.32$ & $20.93 \pm 0.18$ & $0.47 \pm 0.02$ & $29.74 \pm 0.12$ & $0.15 \pm 0.05$\\
        & \multicolumn{1}{l}{B+bar+bl+D1+D2 (geo. fixed)} & $0.34 \pm 0.04$ & $15.28\pm 0.30$ & $21.57\pm 0.03$ & $0.33 \pm 0.01$ & $30.09 \pm 0.09$ & $0.07 \pm0.01$ \\
        & \multicolumn{1}{l}{B+bar+bl+D1+D2 (geo. fixed, $n_{\rm bl}$=1.0)} & $0.36\pm 0.03$ & $15.17 \pm 0.35$ & $21.63 \pm 0.03$ & $0.31 \pm 0.01$ & $29.78 \pm 0.06$ & $0.06 \pm 0.01$ \\
        \hline
        \multirow{3}{*}{barlens} & \multicolumn{1}{l}{B+bar+bl+D1+D2 (free)} & $0.79\pm 0.18$ & $3.69\pm 0.17$ & $19.06\pm 0.04$ & $0.78\pm 0.01$ & $30.87\pm 0.09$ & $0.13\pm 0.02$ \\
        & \multicolumn{1}{l}{B+bar+bl+D1+D2 (geo. fixed)} & $2.06 \pm 0.57$ & $6.50\pm 0.63$(fixed) & $20.37 \pm 0.07$ & $0.79 \pm 0.03$(fixed) & $30.27\pm 0.66$ & $0.18 \pm 0.02$ \\
        & \multicolumn{1}{l}{B+bar+bl+D1+D2 (geo. fixed, $n_{\rm bl}$=1.0)} & $1.00\pm 0.10$(fixed) & $6.50\pm 0.63$(fixed) & $20.29 \pm 0.03$ & $0.79\pm 0.03$(fixed) & $32.94 \pm 0.11$ & $0.14 \pm 0.02$ \\
	\hline
	\end{tabular}
}
\tablefoot{In column (2), `free' stands for freeing all parameters of the barlens; `geo. fixed' stands for fixing the axial ratio and the size of the barlens; `geo. fixed, $n_{\rm bl}$=1.0' stands for fixing the axial ratio, the size and the Sérsic index of the barlens.}
\end{table*}

Model 2 is more complicated than Model 1, as it includes a classical bulge formed early in the galaxy’s evolution, followed by bar formation at a later stage. Figure~\ref{fig:au6_faceon_decomposition} shows the results of 2D image decomposition of the face-on mock image of Model 2. Table~\ref{tab:au6_faceon_2D_params} lists the best-fit parameters of the bulge, bar, and barlens components for different fitting configurations. We first adopt a simple B+D configuration to fit the mock image. The residual map in the left column of Figure~\ref{fig:au6_faceon_decomposition} shows the bar and the ring structure clearly. In the second column, after including a bar, the overall residual decreases significantly. For the bulge, $n$, $r_{\rm e}$ and B/T all decrease by $\sim 50\%$. Additionally, substituting a single exponential disk with a pair of broken exponential disks in the third column further reduces the residuals in the outer disk, resulting in slightly larger bulge parameters. As expected, the residuals continue to decrease as the complexity of the configuration and the number of components increase. However, within the bar region, there is still room for improvement.

After we add a barlens component (i.e., B+bar+bl+D1+D2 with free parameters in the fourth column of Figure~\ref{fig:au6_faceon_decomposition}), $n$, $r_{\rm e}$, and B/T of the bulge all decrease, with B/T reducing from 0.2 to 0.05. However, in this case (fit the barlens with free parameters), the best-fit size of the barlens becomes unrealistically small, even smaller than the minor axis of the bar. This is physically unreasonable and consistent with the conclusion of \citet{Athanassoula2015} that free parameter fitting for the barlens can yield unrealistic values. To address this issue, based on the best-fit ellipse of the visually identified barlens shape from the image shown in Figure~\ref{fig:Au6}, we fix $r_{\rm e,bl}$ to 6.50 arcsec ($\sim0.4a_{\rm bl}$) and the barlens axial ratio to 0.79, denoted as B+bar+bl+D1+D2 (geo. fixed). The result in the fifth column of Figure~\ref{fig:au6_faceon_decomposition} reveals that fixing the barlens size and axial ratio leads to larger bulge parameters compared to the free-fitting scenario. Specifically, B/T increases to 0.1, with $n\sim1.7$ and $r_{\rm e}\sim2.6$ arcsec. However, the barlens Sérsic index increases to $\sim2$, which is larger than the expected values based on Model 1 and other previous works. To further improve the fitting, we perform another test by fixing $n_{\rm bl}$ to 0.5, 1.0, and 1.5, as suggested by previous studies \citep{Athanassoula2015, Laurikainen2018, Erwin2021}. We choose the result for $n_{\rm bl} = 1.0$, since it exhibits the lowest RFF (Figure~\ref{fig:au6_faceon_decomposition}, the sixth column, with a configuration denoted as B+bar+bl+D1+D2 (geo. fixed, $n_{\rm bl} = 1.0$). Results for $n_{\rm bl} = 0.5$ and 1.5 are presented in Appendix~\ref{apx}. Regardless of the value of $n_{\rm bl}$ (free, or fixed to 0.5, 1.0, 1.5), barlens-to-total ratio (bl/T) always remains $\sim0.15$, approximately $50\%$ of the value in the bulgeless Model 1. Additionally, B/T remains low ($\sim$0.1), which is roughly 50\% of the value in the B+bar+D1+D2 configuration. The bulge parameters ($n$ and $r_{\rm e}$) are reduced in all cases compared to the B+bar+D1+D2 configuration, with changes of roughly $\pm0.3$ in $n$ and $\pm0.4$ arcsec in $r_{\rm e}$ for different values of $n_{\rm bl}$.

\begin{figure*}[ht!]
\centering
\includegraphics[width=\textwidth]{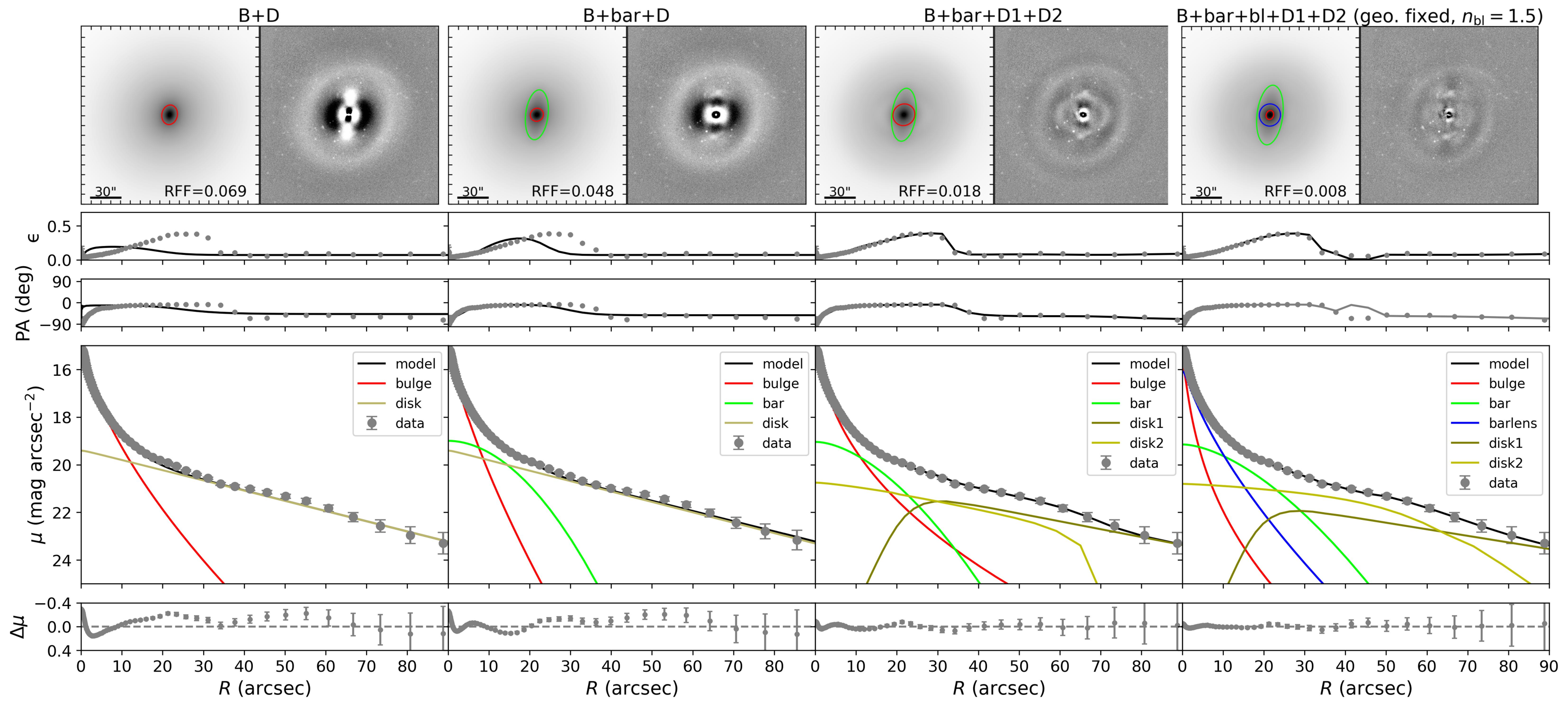}
    \caption{2D image decomposition results of NGC 1533. Other conventions are similar to Figure~\ref{fig:nbody_decomposition}. From left to right, configurations are arranged in the following order: B+D, B+bar+D, B+bar+D1+D2 and B+bar+bl+D1+D2 (fixing the Sérsic index, size and axial ratio of the barlens). }
    \label{fig:NGC1533_decomposition}
\end{figure*}

\begin{table*}[ht!]
	\caption{Best-fit parameters of different components from 2D image decomposition of NGC 1533. The meaning of each column is shown in Table~\ref{tab:nbody_2D_params}. }
	\label{tab:NGC1533_params}
\resizebox{517pt}{!}{
	\begin{tabular}{cccccccc}
		\hline
		Component & Configuration & $n$ & $r_{\rm e}$ & $\mu_{\rm e}$ & b/a & PA & $f/f_{\rm tot}$ \\
             &  &  & (arcsec) & (mag arcsec$^{-2}$) &  & ($^\circ$) &  \\
              (1) & (2) & (3) & (4) & (5) & (6) & (7) & (8) \\ 
		\hline
		\multirow{4}{*}{bulge} & \multicolumn{1}{l}{B+D} & $1.59\pm0.32$ & $5.01\pm 1.31$ & $17.59\pm 0.37$ & $0.79\pm 0.05$ & $-11.03\pm 2.26$ & $0.31\pm 0.03$ \\
        & \multicolumn{1}{l}{B+bar+D} & $1.41\pm 0.12$ & $3.42\pm0.67$ & $17.27\pm0.23$ & $0.95\pm 0.02$ & $-51.56\pm17.36$ & $0.21\pm 0.04$ \\
        & \multicolumn{1}{l}{B+bar+D1+D2} & $2.39\pm 0.54$ & $5.74\pm1.48$ & $18.16\pm0.53$ & $0.94\pm0.01$ & $-32.63\pm11.75$ & $0.33\pm0.04$ \\
        & \multicolumn{1}{l}{B+bar+bl+D1+D2 (geo. fixed, $n_{\rm bl}$=1.5)} & $2.82 \pm 0.15$ & $2.34 \pm 0.15$ & $18.08\pm 0.09$ & $0.80 \pm 0.01$ & $-21.67 \pm 11.36$ & $0.06 \pm 0.01$ \\
        \hline
        \multirow{3}{*}{bar} & \multicolumn{1}{l}{B+bar+D} & $0.52\pm 0.08$ & $12.75\pm1.64$ & $19.77\pm0.44$ & $0.45\pm 0.02$ & $-6.62\pm1.01$ & $0.09\pm 0.03$ \\
        & \multicolumn{1}{l}{B+bar+D1+D2} & $0.51\pm 0.13$ & $13.47\pm0.98$ & $20.00\pm0.17$ & $0.46\pm0.04$ & $-7.26\pm0.21$ & $0.09\pm0.04$ \\
        & \multicolumn{1}{l}{B+bar+bl+D1+D2 (geo. fixed, $n_{\rm bl}$=1.5)} & $0.61\pm 0.04$ & $15.32\pm 0.19$ & $20.11\pm 0.09$ & $0.43\pm 0.01$ & $-7.02\pm 0.22$ & $0.10\pm 0.01$ \\
        \hline
        barlens & \multicolumn{1}{l}{B+bar+bl+D1+D2 (geo. fixed, $n_{\rm bl}$=1.5)} & $1.50\pm 0.10$(fixed) & $5.68\pm 0.84$(fixed) & $18.24 \pm 0.43$ & $0.95\pm 0.04$ (fixed) & $-69.11 \pm 3.95$ & $0.25 \pm 0.02$ \\
		\hline
	\end{tabular}
 }
\tablefoot{In column (2), `geo. fixed, $n_{\rm bl}$=1.5' stands for fixing axial ratio, the size and the Sérsic index of the barlens. }
\end{table*}

RFF is reduced from $\sim$0.04 in B+bar+D1+D2 to $\sim$0.03 in B+bar+bl+D1+D2, confirming that including the barlens component improves the modeling significantly. In principle, the barlens is difficult to model because of the spatial overlap of the barlens with the bar and bulge. Our results confirm that to better model the barlens, it is necessary to fix its geometry parameters and constrain its Sérsic index (also see \citealt{Athanassoula2015}). 

For the parameters of the bar, after using a pair of broken exponential disk to model the ring structure, both $r_{\rm e,bar}$ and bar/T decrease. The Sérsic index of the bar is $\sim$0.4, consistent with a relatively flat light profile, contributing to about 0.1 of the total flux. Nonetheless, in the four configurations using a pair of broken exponential disks, the best-fit bar parameters remain consistent regardless of the changes in the configuration complexity or additional constraints imposed on the barlens parameters. 

Regarding the disk parameters, different from the result in Model 1, when the barlens is included, the parameters of both the inner disk and the outer disk remain roughly unchanged.

\begin{figure*}[ht!]
\includegraphics[width=\textwidth]{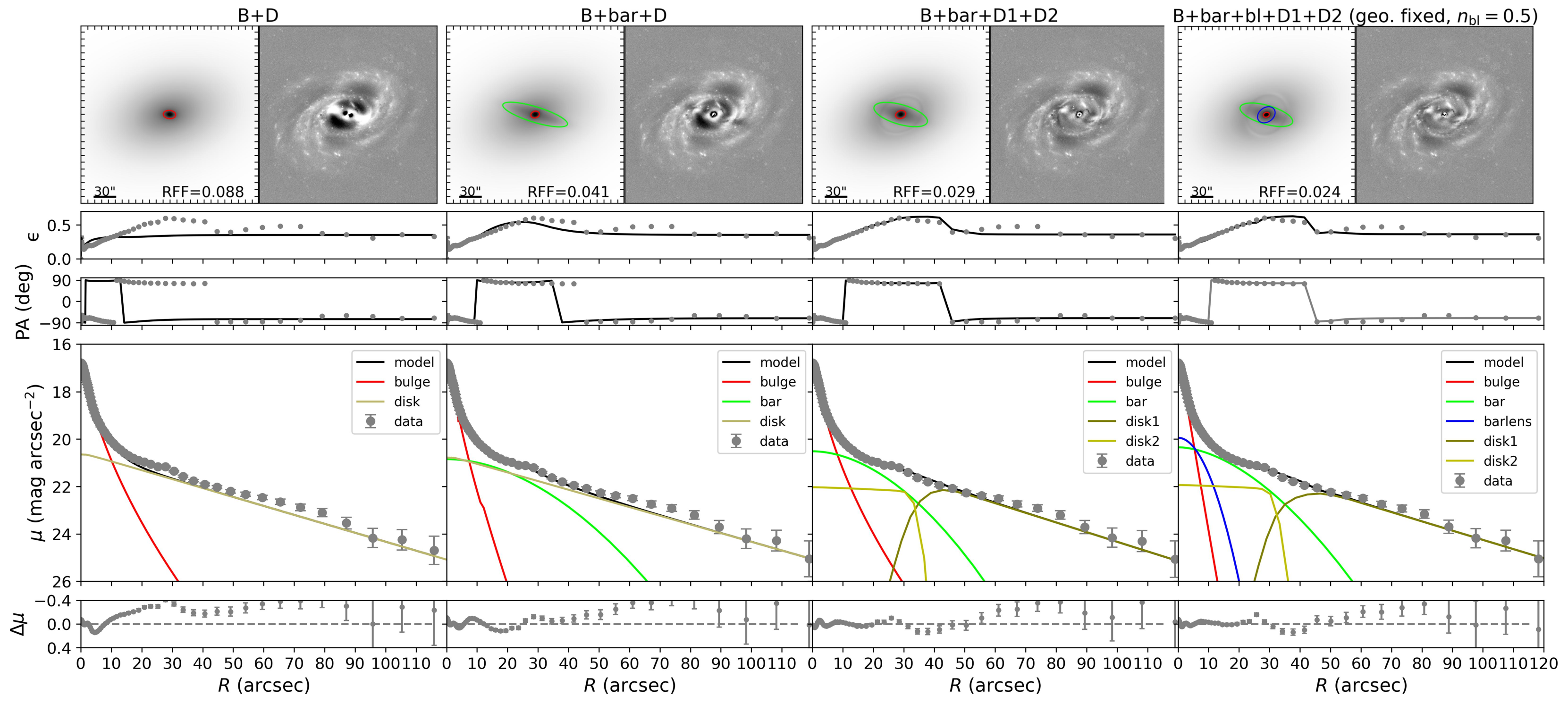}
    \caption{2D image decomposition results of NGC 7329. Other conventions are similar to Figure~\ref{fig:nbody_decomposition}. From left to right, configurations are arranged in the following order: B+D, B+bar+D, B+bar+D1+D2 and B+bar+bl+D1+D2 (fixing the Sérsic index, size and axial ratio of the barlens). }
    \label{fig:NGC7329_decomposition}
\end{figure*}

\begin{table*}[ht!]
	\caption{Best-fit parameters of different components from 2D image decomposition of NGC 7329. The meaning of each column is shown in Table~\ref{tab:nbody_2D_params}. }
	\label{tab:NGC7329_params}
 \resizebox{517pt}{!}{
	\begin{tabular}{cccccccc}
		\hline
		Component & Configuration & $n$ & $r_{\rm e}$ & $\mu_{\rm e}$ & b/a & PA & $f/f_{\rm tot}$ \\
             &  &  & (arcsec) & (mag arcsec$^{-2}$) &  & ($^\circ$) &  \\
              (1) & (2) & (3) & (4) & (5) & (6) & (7) & (8) \\ 
		\hline
		\multirow{4}{*}{bulge} & \multicolumn{1}{l}{B+D} & $1.73\pm 0.46$ & $4.57\pm 1.23$ & $18.95\pm 0.45$ & $0.66\pm 0.07$ & $85.10 \pm 8.38$ & $0.22\pm 0.02$ \\
        & \multicolumn{1}{l}{B+bar+D} & $1.29\pm0.15$ & $3.20\pm0.29$ & $18.48\pm 0.15$ & $0.76\pm 0.01$ & $-73.03\pm 1.75$ & $0.16\pm 0.03$ \\
        & \multicolumn{1}{l}{B+bar+D2+D2} & $1.71\pm 0.12$ & $3.79 \pm 0.26$ & $18.81\pm 0.11$ & $0.75\pm 0.01$ & $-71.19\pm 3.35$ & $0.19\pm 0.05$ \\
        & \multicolumn{1}{l}{B+bar+bl+D1+D2 (geo. fixed, $n_{\rm bl}$=0.5)} & $1.04\pm 0.04$ & $2.31\pm 0.07$ & $18.09\pm 0.04$ & $0.74\pm0.01$ & $ -74.77\pm 2.36$ & $0.11\pm 0.02$ \\
        \hline
        \multirow{3}{*}{bar} & \multicolumn{1}{l}{B+bar+D} & $0.53\pm 0.24$ & $24.68\pm1.70$ & $21.63\pm0.11$ & $0.23\pm 0.01$ & $73.01\pm1.22$ & $0.11\pm 0.04$ \\
        & \multicolumn{1}{l}{B+bar+D1+D2} & $0.54\pm 0.22$ & $24.24\pm1.18$ & $21.28\pm0.05$ & $0.36\pm0.02$ & $74.10\pm0.92$ & $0.15\pm0.06$ \\
        & \multicolumn{1}{l}{B+bar+bl+D1+D2 (geo. fixed, $n_{\rm bl}$=0.5)} & $0.57\pm 0.17$ & $19.98\pm 1.02$ & $21.23\pm 0.07$ & $0.35\pm 0.01$ & $74.31\pm0.51$ & $0.16\pm0.05$ \\
        \hline
        barlens & \multicolumn{1}{l}{B+bar+bl+D1+D2 (geo. fixed, $n_{\rm bl}$=0.5)} & $0.50\pm0.10$ (fixed) & $6.99\pm 0.78$ (fixed) & $20.67\pm 0.06$ & $0.75\pm 0.03$ (fixed) & $-58.11\pm 1.18$ & $0.07\pm 0.02$ \\
		\hline
	\end{tabular}
}
\tablefoot{In column (2), `geo. fixed, $n_{\rm bl}$=0.5' stands for fixing axial ratio, the size of and the Sérsic index of the barlens. }
\end{table*}

\subsection{Applications on the CGS Images}

\subsubsection{NGC 1533}

As shown in the left panel of Figure~\ref{fig:ngc}, NGC 1533, a disk galaxy that is almost face-on, exhibits a short bar, a clear barlens, and an outer ring structure. Following \citet{Gao2017}, we first fit the data with the B+D, B+bar+D and B+bar+D1+D2 configurations. The best-fit parameters for each component are consistent with those in \citet{Gao2017}. Then we add the barlens component and adopt the B+bar+bl+D1+D2 configuration to further refine the fit. The results are shown in Figure~\ref{fig:NGC1533_decomposition} and listed in Table~\ref{tab:NGC1533_params}. From the left to right columns of Figure~\ref{fig:NGC1533_decomposition}, the residual map improves (with a more smoothed residual pattern) due to that we progressively add more structures in the fitting. In B+bar+D1+D2 (the third column in Figure~\ref{fig:NGC1533_decomposition}), a clear ring appears in the central region ($\sim10$ arcsec) of the residual map, suggesting that the central region is not well described by the combination of the bar and the bulge. After adding the barlens component (the fourth column), the ring in the residual map disappears and the RFF decreases from $\sim0.02$ to $\sim0.01$. 

The axial ratio and size of the barlens are determined following the method outlined in Section~\ref{sec:data}. Similar to the decomposition strategy used for Model 2, the Sérsic index of the barlens is fixed to 0.5, 1.0, and 1.5, respectively. For $n_{\rm bl}=1.5$, the corresponding result is shown in the fourth column of Figure~\ref{fig:NGC1533_decomposition} since it has the lowest RFF. In this configuration, the including of the barlens component provides the most significant improvement to the residuals in the central region of the galaxy. bl/T is $\sim$0.25, silghtly larger than that in the mock images of Model 2. After including the barlens, the bulge parameters also change in a manner similar to the simulation results. B/T drops from 0.33 in B+bar+D1+D2 to 0.06 after adding the barlens component in B+bar+bl+D1+D2 (geo. fixed, $n_{\rm bl}$=1.5). Similarly, the effective radius also decrease with the additional barlens introduced. However, the Sérsic index of the bulge increases to $\sim2.8$, different from the results for Model 2. As for the bar component, it has a flat light profile ($n_{\rm bar}\sim0.5$) and accounts for $\sim0.1$ of the total flux in B+bar+D1+D2. After adding the barlens component, the Sérsic index and the effective radius of the bar are slightly increased. For the disk component, with the inclusion of the barlens component, the inner disk scale length continues to increase, while the central surface brightness shows little change. The outer disk is less affected.

Different from the results in Section~\ref{subsec:m2}, fixing $n_{\rm bl}$ to different values significantly influence the parameters of each component in NGC 1533. B/T decreases from 0.18 to 0.06 as $n_{\rm bl}$ increases from 0.5 to 1.5, while the Sérsic index of the bulge increases from $\sim2.0$ to $\sim$2.8 accordingly. bl/T averages around 0.17, with the lowest value of $\sim$0.08 for $n_{\rm bl}=0.5$ and highest value of $\sim$0.25 for $n_{\rm bl}=1.5$. Indeed, accurately modeling the barlens is challenging. In most cases, the bulge parameters would decrease, especially the size and the light fraction B/T. Results for $n_{\rm bl}=0.5$ and 1.0 can be found in Appendix~\ref{apx}.

\subsubsection{NGC 7329}

As shown in the right panel of Figure~\ref{fig:ngc}, the barred galaxy NGC 7329 is observed at a moderate inclination. We follow the fitting strategy in \citet{Gao2017} for NGC 7329 with the B+D, B+bar+D and B+bar+D1+D2 configurations. The fitting results are consistent with those in \citet{Gao2017}. The best-fit model image, residual map, and isophotal ellipse  profiles are shown in Figure~\ref{fig:NGC7329_decomposition} and listed in Table~\ref{tab:NGC7329_params}. From the left to right columns, as more components are included (from B+D to B+bar+D1+D2), the residual map is improved significantly with less prominent residual patterns. However, similar to NGC 1533, the B+bar+D1+D2 configuration (Figure~\ref{fig:NGC7329_decomposition}, the third column) still shows a prominent positive ring feature in the central region ($\sim$5 arcsec) of the residual map, surrounded by a negative ring. This is also evident in the residual profiles, suggesting that an additional component is needed to better model the central region. After adding the barlens component in the B+bar+bl+D1+D2 configuration, the structures in the central region of the residual map, such as the ring in the third column, become less prominent (Figure~\ref{fig:NGC7329_decomposition}, the sixth column). 

Again, the barlens parameters are determined following the method in Section~\ref{sec:data} and are fixed in the fitting process. Similarly, $n_{\rm bl}$ is fixed to 0.5, 1.0 and 1.5. We find that using $n_{\rm bl}=0.5$ produces the flattest residual profile and the smallest RFF among the three cases and is therefore shown in Figure~\ref{fig:NGC7329_decomposition} and Table~\ref{tab:NGC7329_params}. The bulge parameters remain unaffected by the choice of the barlens Sérsic index. The Sérsic index of the bulge ranges between 1.0 to 1.5 for different $n_{\rm bl}$ and B/T remains $\sim0.11$. Additionally, changing $n_{\rm bl}$ does not significantly affect bl/T, which remains $\sim0.07$. Results for $n_{\rm bl}=1.0$ and 1.5 are presented in Appendix~\ref{apx}. 

bl/T equals to 0.07 when $n_{\rm bl}$ is fixed to 0.5, relatively smaller than that for NGC 1533. By comparing the bulge parameters between the B+bar+D1+D2 and the B+bar+bl+D1+D2, we find that including the barlens will slightly decrease the bulge size ($r_{\rm e}$ of bulge varies from 3.79 to 2.31), flatten the bulge light profile ($n$ of bulge varies from 1.71 to 1.04), and decrease the bulge light fraction (B/T from 0.19 to 0.11). These variations are consistent with the results observed in the simulated mock images and in NGC 1533.

\section{Discussion} \label{sec:disc}

\subsection{Comparison with Previous Works}
 
Previous studies have explored methods to model the barlens in 2D image decomposition. \citet{Athanassoula2015} used GALFIT to analyze mock images from $N$-body simulations of barred galaxies and compared them with observations from the Near-IR S0 Galaxy Survey \citep[NIRS0S,][]{Laurikainen2011} and Spitzer Survey of Stellar Structure in Galaxies \citep[S$^4$G,][]{Sheth2010} samples. They employed generalized elliptical isophotes to fit all components \citep{Peng2010}, with the Sérsic function applied to model the light profile of the bulge and barlens, and an exponential profile for the disk. The bar was modeled using a modified Ferrer function. Although an iterative process was adopted to refine the fit, their modeling strategy only considers the B+bar+bl+D configuration, rather than incrementally increasing the model complexity. As a result, it was not possible to assess the specific impact of the additional barlens component on the bulge parameters. The barlens size and axial ratio in their work were also fixed based on the visually identified barlens shape. However, in their method, the Sérsic index was left free, different from our method. In addition, they also fix the bar length from visual estimation. The ring structure in the disk was also neglected in their fitting. 

Later, \citet{Laurikainen2018} performed 2D image decomposition for 46 barlens galaxies from the CALIFA survey using GALFIT. Similar to \citet{Athanassoula2015}, they also employed generalized elliptical isophotes for all components and used the same fitting functions, except for the disk, which was modeled with a Sérsic function. Similar to our fitting procedure, they began with a single component fitting and progressively increased the model complexity, but the ring structure was ignored. They initially visually estimated the size and axial ratio of the barlens, which were then fixed in subsequent fitting process. The size of the barlens was only fixed when necessary, while the axial ratio was kept fixed throughout. The galaxy center and disk orientation were also fixed during their fitting process. However, different from our method, their Sérsic index of the barlens was left free, similar to \citet{Athanassoula2015}. This may partly explain the presence of some outliers (with large Sérsic index) in their results.

Recently, \citet{Erwin2021} performed 2D image decomposition on two strongly barred galaxies, NGC 4608 and NGC 4643, using IMFIT \citep{Erwin2015}. In their work, the barlens, referred to as the B/P bulge, was modeled using generalized elliptical isophotes. The ring structure was modeled with an additional Gaussian component, while the surface brightness distribution of the outer flat bar was modeled with the default broken-exponential profile. They started with the B+D configuration and progressively increased model complexity, and all parameters in the fitting process were set free. For the two galaxies in \citet{Erwin2021}, and the tests mentioned in their appendix, the barlens Sérsic index ranges from 0.5 to 1.1, similar to the values obtained here. 

Although the fitting strategies in these previous works differ from ours, their results are generally consistent with our findings. \citet{Athanassoula2015} demonstrated that the barlens can produce a central peak in the galaxy light profile with a Sérsic index less than 2, which aligns with our results for the mock images. The measured bl/T ratios in their models range from 0.13 to 0.43, with an average value of $\sim$0.2, similar to our results. In \citet{Laurikainen2018}, the Sérsic index of the barlens also agrees with our findings. On average, the barlens in their sample has $n_{\rm bl} \sim 1$ and bl/T is $\sim10-30\%$. However, there are four galaxies with $2.0 \leq n_{\rm bl} \leq 2.5$. These relatively large Sérsic indexes of the barlens highlight the necessity of reasonable constraints on $n_{\rm bl}$ in the image decomposition
  
For the bulge component, comparing B/T between different fitting configurations, \citet{Laurikainen2018} found that B/T decreases from $\sim$0.3 (B+D), to $\sim$0.15 (B+bar+D), and to $\sim$0.06 (B+bar+bl+D), which is consistent with our results. For Model 2, we observed a similar trend, with B/T decreasing from $\sim$0.45 (B+D), to $\sim$0.25 (B+bar+D1+D2), and to $\sim$0.10 (B+bar+bl+D1+D2). Similarly, both the bulge Sérsic index and the effective radius decrease after adding the barlens component. In \citet{Erwin2021}, compared to B+D, considering a barlens component leads to a decrease in B/T (from $\sim0.5-0.7$ to $\sim0.1$) and the Sérsic index (from $\sim3$ to $\sim2$). Moreover, \citet{Erwin2021} utilized the IFU data to derive the stellar kinematics of the galaxies and assess the reliability of the bulge type classification based on the Sérsic index obtained from the 2D image decomposition with a barlens component. The IFU kinematics confirmed the presence of a classical bulge ($n \sim 2.2$) in NGC 4608, a nuclear disk (with a broken exponential distribution) with a compact classical bulge ($n \sim 1.6$) in NGC 4643, which were identified in the 2D image decomposition. 

\subsection{The Effect of Disk Inclination}

\begin{figure}[b!]
\includegraphics[width=\columnwidth]{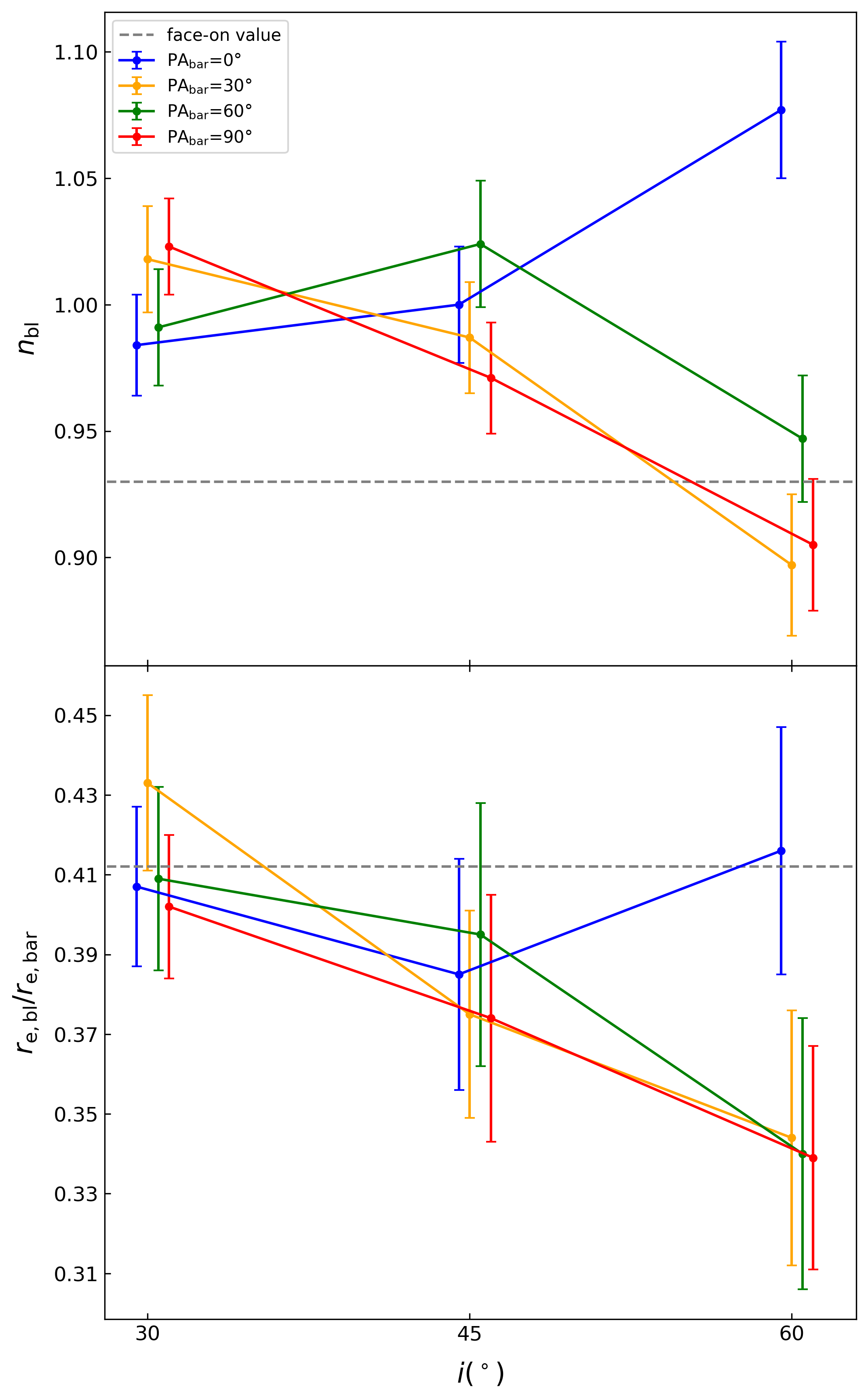}
    \caption{Sérsic index (top panel) and relative size (bottome panel) of the barlens for Model 1 at different disk inclinations and bar position angles. The horizontal dashed line in each panel represents the best-fit $n_{\rm bl}$, or $r_{\rm e,bl}/r_{\rm e.bar}$ from the face-on mock image of Model 1. As the disk inclination angle increases, both the Sérsic index and the relative barlens size decrease. However, for the model with PA$_{\rm bar}$=$0^\circ$, where the major axis of the bar aligns with the minor axis of the projected disk, $n_{\rm bl}$ and $r_{\rm e,bl}/r_{\rm e.bar}$ increase at $i=60^\circ$. }
    \label{fig:bl_i}
\end{figure}

\begin{figure*}[ht]
\includegraphics[width=\textwidth]{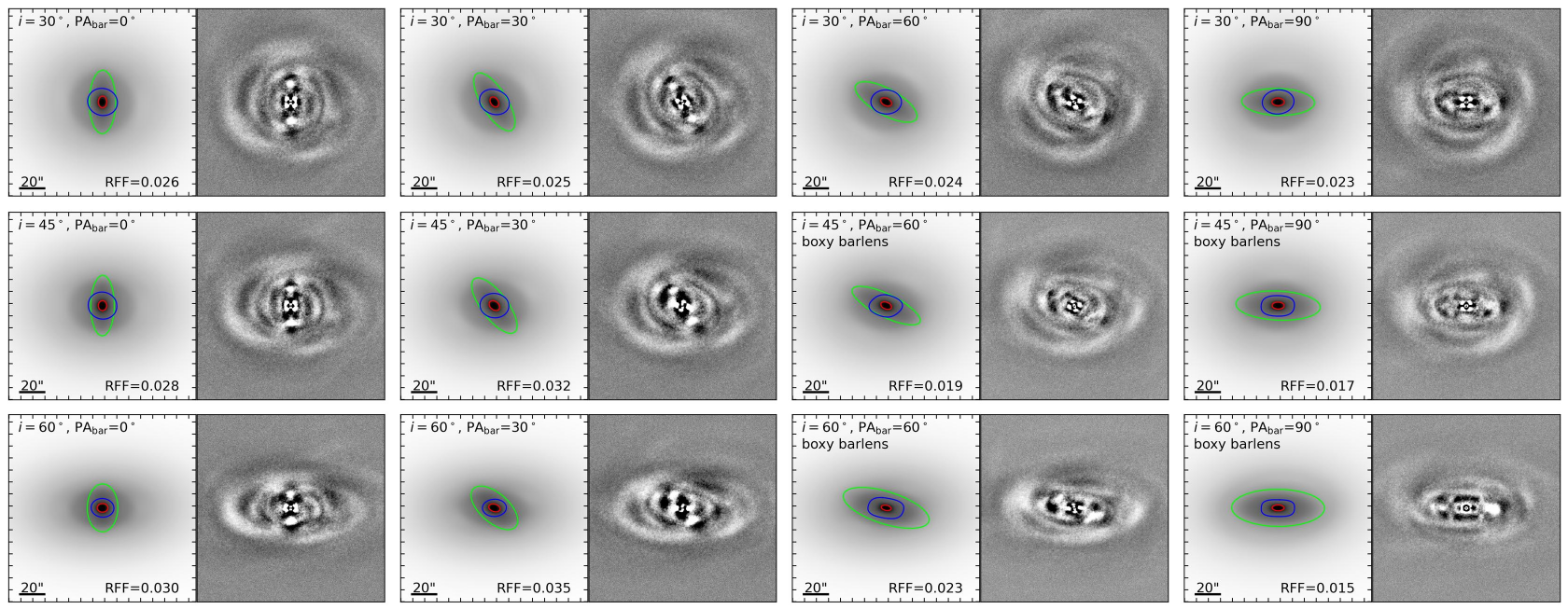}
    \caption{Best-fit B+bar+bl+D1+D2(fixed) model images and residual maps of the 2D image decomposition for Model 2 at different inclinations and PA$_{\rm bar}$. From top to bottom rows, the inclination angle increases from 30$^\circ$ to 60$^\circ$. From left to right columns, PA increases from 0$^\circ$ to 90$^\circ$. Note that the boxy isothote is used for the barlens at large inclination angle ($i=45^\circ$ and $60^\circ$, and PA$_{\rm bar}$=60$^\circ$ and $90^\circ$). }
    \label{fig:au6_i_model}
\end{figure*}

\begin{figure*}[ht]
\includegraphics[width=\textwidth]{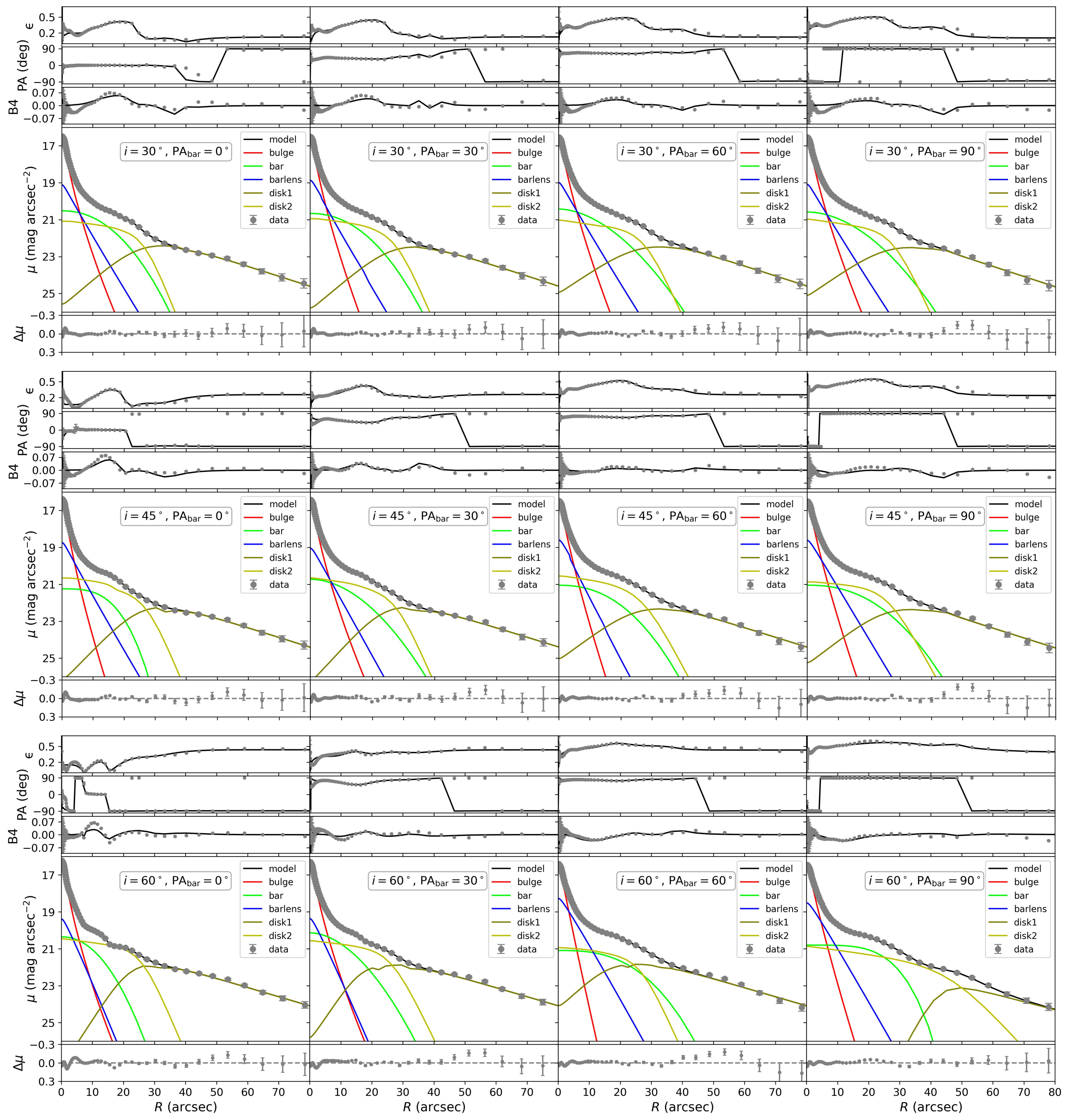}
    \caption{Isophotal analysis showing the ellipticity, PA, $B_4$ (negative for boxyness, positive for diskyness), surface brightness and residual profile of the data images and the corresponding best-fit model images in Figure~\ref{fig:au6_i_model}. The top, middle, and bottom rows show the results at $i=30^\circ$, $45^\circ$, and $60^\circ$, respectively. From left to right, PA$_{\rm bar}$ increases from 0$^\circ$ to 90$^\circ$. }
    \label{fig:au6_sbp}
\end{figure*}

We now investigate the influence of disk inclination ($i$) on the modeling of the barlens and the bulge. For Model 1 and Model 2, mock images are generated with $i=30^\circ$, $45^\circ$ and $60^\circ$, and PA$_{\rm bar}$=$0^\circ$, $30^\circ$, $60^\circ$ and $90^\circ$. The image decomposition strategy described in Section~\ref{sec:data} is applied to these mock images. First, we examine Model 1 at different $i$ and PA$_{\rm bar}$ to study how the barlens properties change with viewing angle. We then analyze Model 2 to understand how the bulge parameters vary with different inclination angles.

\begin{figure}[ht]
\centering
\includegraphics[width=\columnwidth]{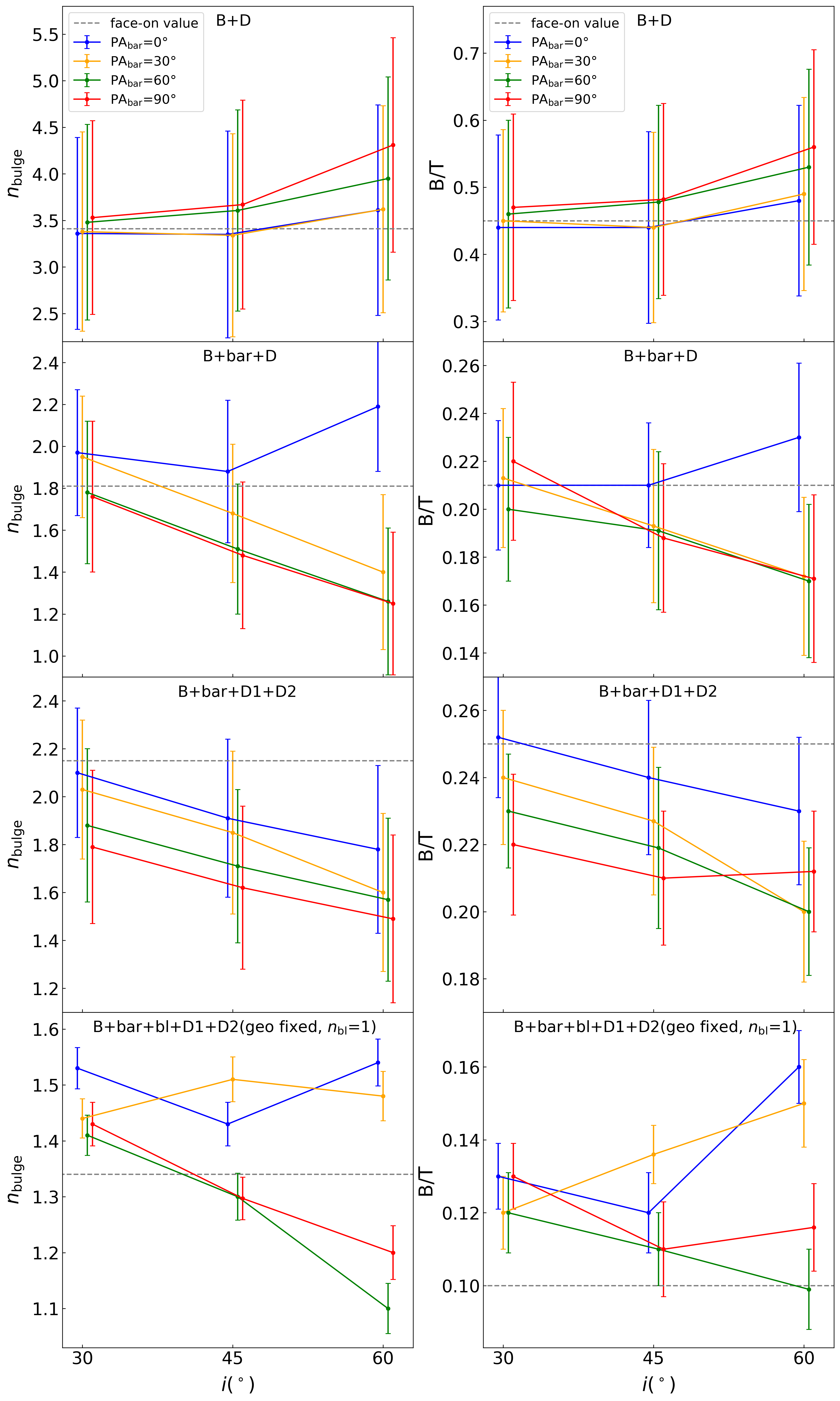}
    \caption{Sérsic index (left column) and B/T (right column) of the bulge of Model 2 at different inclinations and PA$_{\rm bar}$ values for different fitting configurations. From top to bottom, the complexity of the fitting configuration increases. }
    \label{fig:au6_i_analyze}
\end{figure}

\begin{figure}[ht]
\centering
\includegraphics[width=\columnwidth]{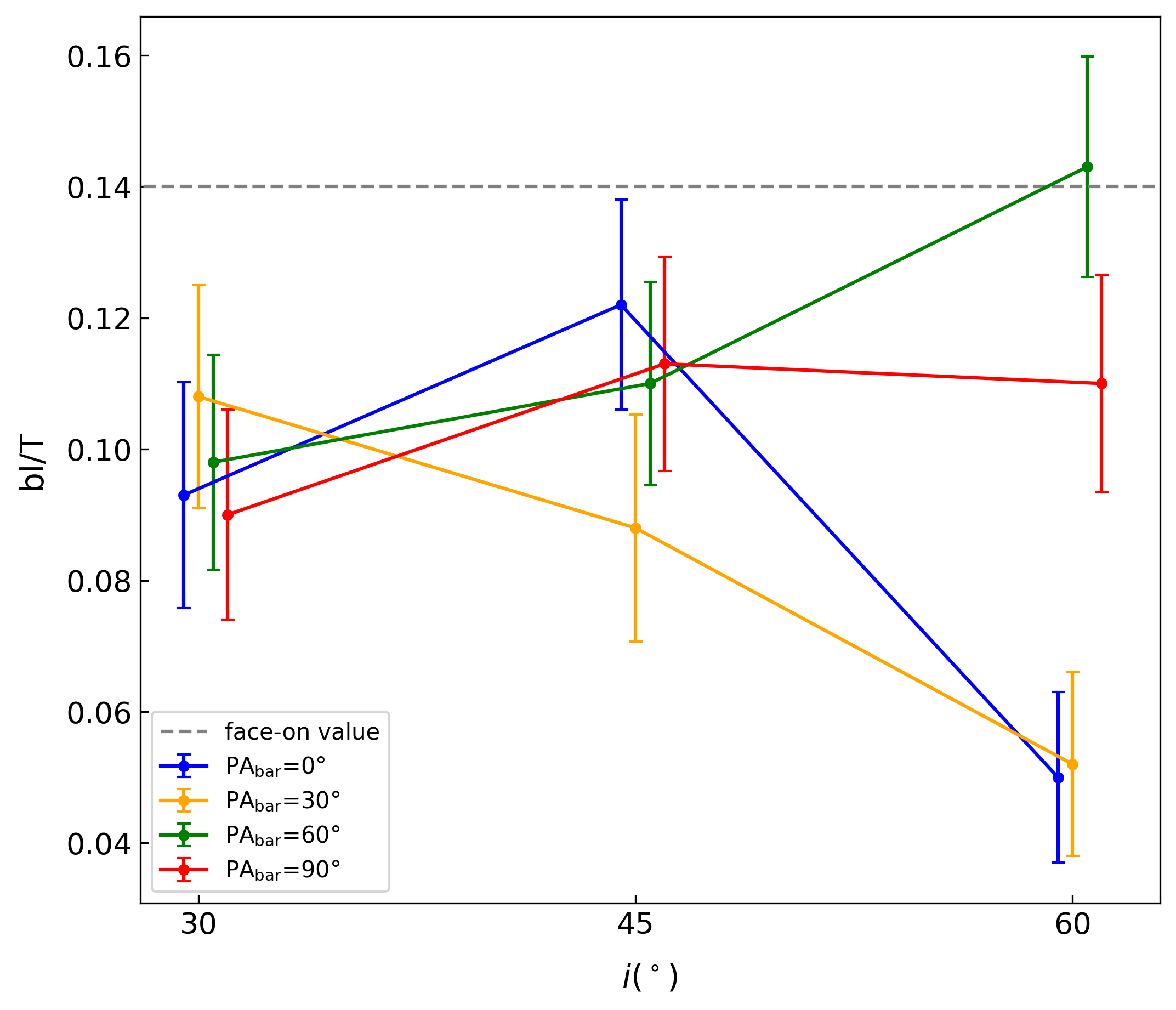}
    \caption{Light fraction of the barlens (bl/T) of Model 2 in the B+bar+bl+D1+D2 (geo. fixed, $n_{\rm bl}$=1.0) configuration at different inclinations and PA$_{\rm bar}$ values. }
    \label{fig:au6_i_bl}
\end{figure}

Figure~\ref{fig:bl_i} shows the variation of the Sérsic index and the relative size of the barlens (with respect to the bar effective radius $r_{\rm e,bl}/r_{\rm e,bar}$) for Model 1. The top panel suggests that the best-fit surface brightness profile of the barlens is only slightly affected by the disk inclination, with $n_{\rm bl}$ decreasing from $\sim1$ to $\sim0.9$. In the bottom panel, the relative size of the barlens ($r_{\rm e,bl}/r_{\rm e,bar}$) also shows minimal variation with inclination, decreasing slightly from $\sim0.41$ to $\sim0.35$ as the inclination angle increases. In addition, for a given inclination angle, the Sérsic index and the relative barlens size across different PA$_{\rm bar}$ values are roughly similar, suggesting that the bar position angle is less important here. However, for the model with PA$_{\rm bar}$=$0^\circ$ at $i=60^\circ$, the best-fit barlens parameters deviate from  those of other PA$_{\rm bar}$ values. In this case, the major axis of the bar is aligned with the minor axis of the projected disk, producing a less elongated bar structure. This alignment results in a serious degeneracy between different components in the central region, leading to a higher Sérsic index and a larger relative size of the barlens than that for other PA$_{\rm bar}$.

Next, we analyze the mock images of Model 2 at different inclination angles with the same fitting strategy as Section~\ref{subsubsec:strategy}. We fix the Sérsic index of the barlens to 1.0 based on the previous results. We also test with $n_{\rm bl}$ fixed to 0.5 and 1.5, and find consistent results with those for $n_{\rm bl}=1.0$. The size and axial ratio of the barlens are determined and fixed using the same method as in Section~\ref{subsec:m2} for the face-on view. The best-fit configurations of B+bar+bl+D1+D2 (geo. fixed, $n_{\rm bl}$=1.0) and the corresponding residual maps are shown in Figure~\ref{fig:au6_i_model}, with corresponding isophote profiles (ellipticity, PA, and surface brightness) in Figure~\ref{fig:au6_sbp}. These configurations provide a good representation of both the images and the surface brightness profiles at different inclination angles.

Figure~\ref{fig:au6_i_analyze} presents a comparison of the bulge Sérsic index and B/T for different fitting configurations for Model 2 at various inclinations. For the bulge component, both B/T and $n$ generally decrease with increasing the complexity of the fitting configurations for different inclinations, similar to the results in Section~\ref{subsec:m2} for the face-on mock images of Model 2. Our results show that B/T and $n$ of the bulge component are roughly consistent across different inclination angles and PA$_{\rm bar}$ values. At all inclination angles, B/T is $\sim$0.5 with the B+D configuration. Once the bar is introduced in the B+bar+D configuration, B/T drops significantly to $\sim$0.2. Substituting a simple exponential disk with a pair of broken exponential disks leads to a slight increase in B/T. The inclusion of the barlens further decreases B/T to $\sim$0.1. The Sérsic index $n$ of the bulge also drops from $\sim$4 in the B+D configuration to $\sim$1.4 when both the bar and barlens are included. In conclusion, the bulge parameters are not strongly affected by the inclination, while the inclusion of the bar and barlens components significantly reduces both B/T and $n$ values for the bulge.

The comparison of bl/T between different inclinations and PA$_{\rm bar}$ values are shown in Figure~\ref{fig:au6_i_bl}. On average, bl/T is $\sim$0.11. As the inclination angle increases, the difference of bl/T between different PA$_{\rm bar}$ values increases. At higher inclination angle ($i=60^\circ$) with PA$_{\rm bar}$=$0^\circ$ and $30^\circ$, bl/T decreases. This is likely due to that  distinguishing between different substructures becomes more challenging when the bar aligns with the minor axis of the projected disk at higher inclinations. \citet{Athanassoula2015} also examined the inclination effect on bl/T and found a larger variation between different PA$_{\rm bar}$ values at $i=60^\circ$ (ranging between $\sim$0.05 and $\sim$0.12) than $i=30^\circ$ (ranging between $\sim$0.11 and $\sim$0.13), which is consistent with our results. 

At even larger inclinations ($i > 60^\circ$), determining the barlens parameters becomes increasingly challenging, primarily due to the overlap between the bulge, bar, barlens, and disk. This overlap leads to a more severe parameter degeneracy, resulting in larger uncertainties in the structural parameters. Additionally, the projection effect can change the intrinsic light profiles and isophotal shapes of the different components compared to the face-on case.

Based on the measurements conducted, we can infer a possible correction for the bulge properties of a moderately inclined galaxy ($i\leq60^\circ$) to recover their intrinsic values in the face-on case. For simple fitting configurations (B+D and B+bar+D), $i$ decides the amount of deviation and PA$_{\rm bar}$ influences the direction of the deviation (whether the values are over or underestimated). The Sérsic index and B/T deviates by 10–30\% from the face-on values, with the deviation increasing with $i$. It is higher than the face-on value at low PA$_{\rm bar}$ and lower at high PA$_{\rm bar}$. For the B+bar+D1+D2 configuration, inclination still introduces systematic deviations, but PA$_{\rm bar}$ has a comparatively weaker effect on these parameters. Once again, the Sérsic index and B/T exhibits a 10-30\% deviation from the face-on value, with the deviation increasing with $i$. After including the barlens component in B+bar+bl+D1+D2 (geo. fixed, $n_{\rm bl}$=1.0), at low inclinations ($i\leq30^\circ$), the Sérsic index and B/T is systemically higher for $\sim15\%$ across all PA$_{\rm bar}$ than the face-on value. At higher inclinations, the bulge properties remain $\sim15\%$ higher than the face-on value at lower PA$_{\rm bar}$; for higher PA$_{\rm bar}$, the Sérsic index is 10-20\% lower than the face-on value, with the deviation increasing with $i$, while B/T remains 10\% higher than the face-on value. 

We also tested image decomposition of the edge-on mock images ($i = 90^\circ$) for both Model 1 and Model 2. Due to the severe degeneracy of these components and the complex light profiles and irregular isophotal shapes of the disk, bar, and barlens in the edge-on view, the overall residuals are too large for the fittings to be considered acceptable. We decided not to include these results in our analysis. The 2D image decomposition of the edge-on galaxies is difficult and challenging, which should be investigated in more detail in the future work.

\subsection{Boxy/Peanut Bulge Modeling}

\begin{figure}[ht]
\includegraphics[width=\columnwidth]{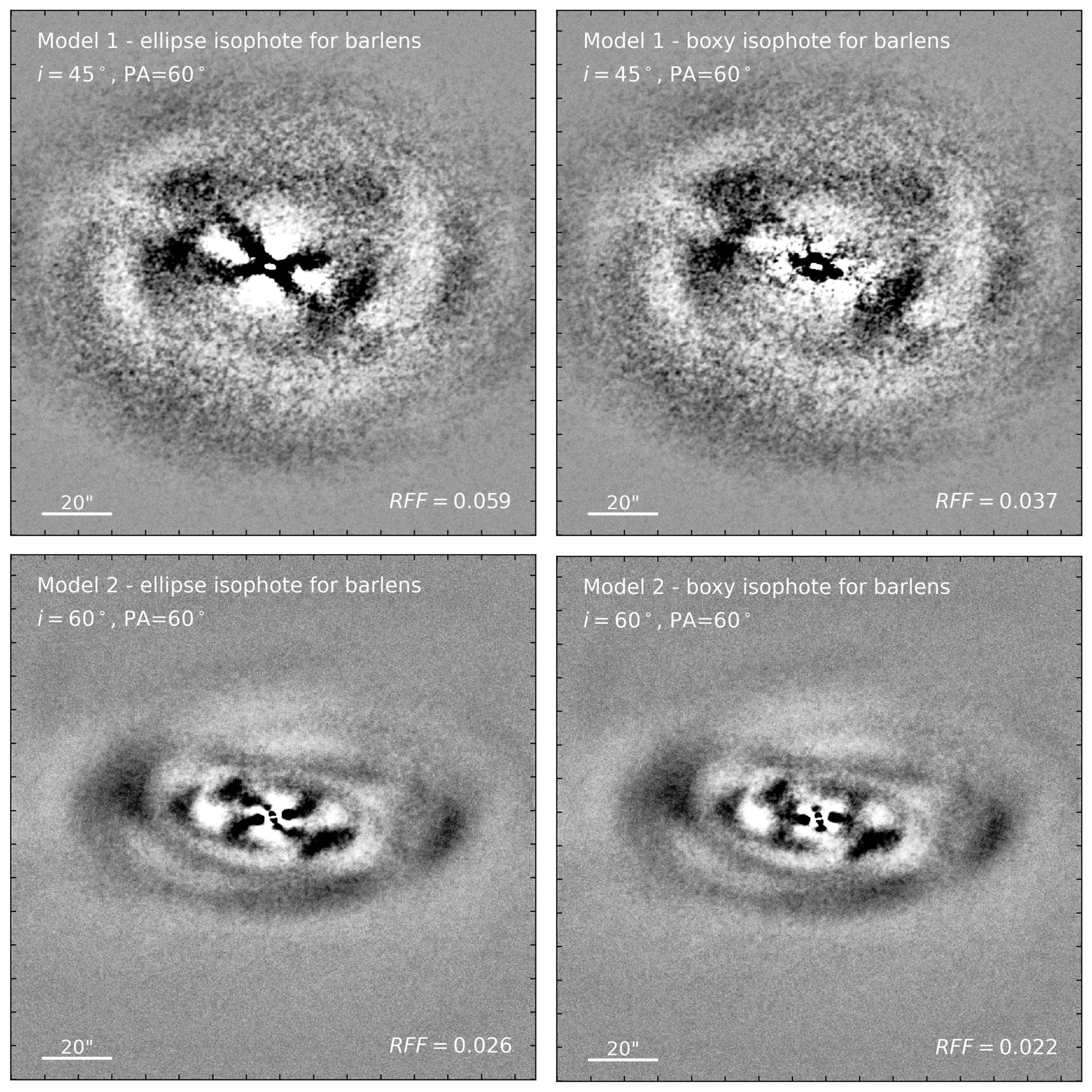}
    \caption{Comparison of the residual maps of the configurations with the ellipse isophote (left) and the boxy isophote for the barlens component (right). The top and bottom rows show the results for Model 1 and Model 2, respectively. }
    \label{fig:residual}
\end{figure}

As mentioned in Section~\ref{sec:data}, we use the Fourier mode to model the barlens that becomes the B/P bulge for PA$_{\rm bar}$=$60^\circ$ and $90^\circ$ at high inclinations ($i=45^\circ$ and $60^\circ$) for both Model 1 and Model 2 (as shown in Figure~\ref{fig:au6_i_model}). At these inclinations, visually determining the axial ratio of the B/P bulge is challenging. After performing the initial free parameter fitting for the highly inclined galaxies, the barlens axial ratio is manually adjusted to better match the B/P bulge shape. The residual maps for the configurations with boxy isophotes of the barlens are compared to those with elliptical isophotes in Figure~\ref{fig:residual}. As expected, using an elliptical isophote results in a distinct X-shaped feature in the residuals, which is properly modeled by the boxy isophote. For both Model 1 and Model 2, compared to previous results for the barlens, using boxy isophotes reduce RFF in the identified barlens region by $\sim$30\% on average (from $\sim0.060$ to $\sim0.037$ for Model 1 and from $\sim0.025$ to $\sim0.018$ for Model 2), confirming that our method of modeling the barlens with boxy isophotes at large inclinations is reasonable. Note that for both Model 1 and Model 2, the parameters of the bulge, bar, and barlens remain virtually unchanged before and after adopting the boxy isophote for the barlens.

\subsection{BIC Test}

The existence of the barlens component has been validated in both simulations and observations \citep{Laurikainen2014, Athanassoula2015, Laurikainen2017, Li2017, Erwin2017}. In the 2D image decomposition, adding extra components can always improve the fitting. However, it also increases the model complexity to result in potential overfitting issue. It is important to assess whether an additional component is truly necessary. 

In the literature, several methods have been proposed to quantify the necessity of adding an additional component. For example, \citet{Erwin2021} used the Akaike information criterion (AIC; \citealt{Akaike1974}) to quantify the improvement in the results after modeling the barlens in the 2D image decomposition. They observed a dramatic decrease in AIC (larger than 1000) after modeling the barlens, highlighting the reliability of their fitting. 

The Bayesian Information Criterion (BIC) is useful to evaluate whether a new component should be added in the fitting \citep{Schwarz1978}. This criterion evaluates the likelihood of a model while applying a penalty for the number of free parameters in the fit. Compared to AIC, BIC imposes a more stringent penalty for an increase in the number of free parameters. According to \citet{Head2014, Mendez2018}, BIC statistic can be written as:
\begin{equation}
    {\rm BIC}=\frac{\chi^2}{A_{\rm psf}}+k \ln(\frac{N}{A_{\rm psf}}), 
	\label{eq:bic}
\end{equation}
where
\begin{equation}
    \chi^2=\sum_{i=1}^{N} \frac{[d_i-f_i]^2}{\sigma^2_i}.
	\label{eq:deviation}
\end{equation}
$k$ is the number of free parameters and $N$ is the pixel number of the image. $A_{\rm psf}$ is the size area of the PSF at the full width at half maximum (FWHM). $d_i$ is the flux of the $i$-th pixel in the data image, and $f_i$ is the flux of the $i$-th pixel in the generated model. $\sigma^2_i=d_i+S_i$, where $S_i$ is the noise. Note that the BIC statistic cannot be the sole criterion for determining the number of components used in the fitting, as its effectiveness is limited by the complexity of real galaxy structures \citep{Argyle2018} and may lead to inconsistent results compared to visual inspection. The BIC test should be complemented by other strategies when determining the necessity of including a barlens in the galaxy decomposition, such as visual inspection or isophotal ellipse fitting. The parameters from 2D image decomposition, such as B/T, should be also taken into account when employing the BIC test \citep{Nedkova2024, Bellstedt2024}. 

\begin{table}
\caption{BIC for different configurations in different galaxies. The columns show: (1) the used galaxy; (2) the configuration; (3) BIC. }
\label{tab:BIC}
\begin{tabular}{ccc}
\hline
Galaxy & Configuration & BIC
 \\
\hline
\multirow{3}{*}{Model 1} & bar+D & 26697 \\
& bar+D1+D2 & 9700 \\
& bl+bar+D1+D2 (geo. fixed) & 8550 \\
\hline
\multirow{4}{*}{Model 2} & B+D & 4588 \\
& B+bar+D & 4214 \\
& B+bar+D1+D2 & 3597 \\
& B+bar+bl+D1+D2 (geo. fixed, $n_{\rm bl}$=1.0) & 3599 \\
\hline
\multirow{4}{*}{NGC 1533} & B+D & 21320 \\
& B+bar+D & 21053 \\
& B+bar+D1+D2 & 19091 \\
& B+bar+bl+D1+D2 (geo. fixed, $n_{\rm bl}$=1.5) & 17198 \\
\hline
\multirow{4}{*}{NGC 7329} & B+D & 83842 \\
& B+bar+D & 84101 \\
& B+bar+D1+D2 & 77800 \\
& B+bar+bl+D1+D2 (geo. fixed, $n_{\rm bl}$=0.5) & 77689 \\
\hline
\end{tabular}
\end{table}

Table~\ref{tab:BIC} shows the values of BIC for different fitting configurations in different galaxies. When counting the number of free parameters used in calculating BIC, the fixed parameters of the barlens are also included. Configurations with lower BIC values are considered superior. Additionally, it is important to examine the BIC variation between different fitting configurations when new components are included. Typically, a decrease in BIC of more than $\sim10$ is considered sufficient to indicate that a more complex configuration is preferred over a simpler one \citep{Argyle2018}. As presented in Table~\ref{tab:BIC}, adding a bar component and adopting a pair of broken exponential disks can decrease BIC considerably (except for NGC 7329 with B+bar+D). For Model 1, we can see a clear reduction of the BIC by modeling the barlens when comparing the bar+D1+D2 and the bl+bar+D1+D2 configurations. For the two CGS galaxies, we can also see an appreciable decrease in BIC after adding the barlens component, demonstrating the necessity of modeling the barlens. However, for Model 2, the barlens component hardly affects BIC. This is expected, as in Model 2 the residual pattern in the central region of B+bar+bl+D1+D2 (geo. fixed, $n_{\rm bl}$=1.0) does not change significantly compared to B+bar+D1+D2. This is different from the results in the CGS real galaxies, where the residuals in the central region become much less significant. The difference may indicate that the barlens in Model 2 exhibits stronger degeneracy with neighboring structures. Therefore, including a barlens component does not significantly improve the BIC. Model 2 could have more significant light concentration in the central region, to result in higher weights during 2D image decomposition. In this case, it is more important to combine the isophotal shapes and the residual surface brightness profiles to justify the need for a barlens component for Model 2 (RFF decreases from 0.035 to 0.03 after the barlens is added).

\subsection{Potential Impacts and Future Prospects}

We have shown that the inclusion of the barlens component in 2D decomposition could potentially affect the bulge parameters, including the bulge Sérsic index, effective radius, and B/T. These changes may have non-negligible implications on the well-established scaling relations of the bulges in the barred galaxies.

For example, the Kormendy relation \citep{Kormendy1977} links the effective radius and surface brightness of spheroidal systems, with the classical bulges following the relation defined by ellipticals \citep{Kormendy1985, Kormendy1987, Bender1992}, and pseudobulges lying below as low surface brightness outliers \citep{Carollo1999, Gadotti2009, Fisher2010, Gao2020}. After considering the barlens, the effective radius and surface brightness of the bulge are affected to change its position within the Kormendy relation. The bulge classification (pseudo or classical) could also be affected after including the barlens, since the Sérsic index of the bulge also changes. The modeling of the barlens may also affect the bulge mass–size relation. However, previous studies of composite bulges suggest that the overall impact of different decomposition configurations on this relation is relatively modest \citep{Erwin2015C}.

Another possible influence is on the B/T distribution along the Hubble sequence. Since the inclusion of the barlens component can significantly reduce B/T, the B/T difference between the early-type and late-type galaxies may become larger depending on the fraction of the barred galaxies in each morphological types. In addition, B/T in the barred galaxies could be smaller than that in the unbarred galaxies, whereas previous studies have found them to be roughly similar \citep{Gao2019}. 

The impact of these above mentioned effects on the scaling relations of the bulges remains unclear and will be examined in our future work with larger observational datasets. Given that our current work is primarily focused on evaluating barlens modeling techniques and establishing the necessity of including the barlens component in the 2D image decomposition, we do not attempt to draw broad conclusions regarding the behavior of these scaling relations.

\section{Conclusions} \label{sec:conc}

In this paper, we explore the optimal method for modeling the barlens and the impact of this method on the bulge parameters in galaxy 2D image decomposition, using a pure barred disk $N$-body simulation (Model 1) and a Milky Way-like disk galaxy from the Auriga simulation (Model 2). We also apply our decomposition technique to two selected galaxies from the CGS and examine the effect of inclination on the modeling of both the bulge and barlens. Moreover, to assess the necessity of including the barlens, we conduct BIC tests on the decomposition results. While our sample is limited, this study serves to demonstrate an effective strategy for modeling the barlens in 2D image decomposition, highlighting the importance of accurately accounting for the barlens component to obtain reliable bulge and disk parameters. A more comprehensive statistical analysis based on a larger sample of CGS galaxies will be presented in future work.

The main results are as follows: 
\begin{enumerate}
    \item The barlens has a near-exponential surface brightness profile with $n_{\rm bl} \sim 0.9$ (based on the results in Model 1). In realistic galaxies with a distinct bulge component (as seen in Model 2 and the two observed galaxies), the barlens accounts for 10-20\% of the total galaxy light.
    \item In 2D image decomposition, modeling the barlens component helps smooth out the central residual patterns, reduces the RFF, and significantly decrease the BIC. To correctly model the barlens, it is essential to constrain the size, axial ratio, and Sérsic index of the barlens. Without these constraints, the fitting may yield unreasonable parameter. The size and axial ratio can be constrained based on the visually identified outer edge of the barlens in the galaxy image. The Sérsic index of the barlens, according to results from Model 1 and previous studies, should be constrained between 0.5 and 1.5. 
    \item Ignoring the barlens component in 2D image decomposition can significantly affect the bulge measurements. B/T can be overestimated by approximately 30-60\%. When the barlens is correctly modeled, only about 10-20\% of the total galaxy light is attributed to the bulge component. Other bulge parameters, such as the Sérsic index and effective radius, are also reduced after modeling the barlens component. 
    \item Both the bulge and barlens parameters are only weakly affected by the inclination of the galaxy or the position angle of the bar. For the barlens component, the Sérsic index of the barlens, $n_{\rm bl}$, shows a slight decrease from $\sim$1 to $\sim$0.9, and the relative size of the barlens, $r_{\rm e,bl} / r_{\rm e,bar}$, decreases from $\sim$0.41 to $\sim$0.35 as the inclination angle increases. The influence of the PA$_{\rm bar}$ is even less minimal. For the bulge component, both B/T and the Sérsic index decrease with increasing model complexity for different inclinations. These parameters remain relatively consistent across different inclination angles and PA$_{\rm bar}$ values. However, as the galaxy becomes more inclined, the parameters deviate from the face-on situation significantly, and the spatial overlap of different structures leads to higher uncertainties in the parameter estimates.
    \item When the galaxy has a high inclination angle and a large bar position angle, the barlens should be modeled with boxy isophotes to better capture its shape and improve the central residuals. 
\end{enumerate}

\begin{acknowledgements}
We thank the referee for the valuable suggestions that helped improve our work. We thank Luis Ho, Martin Bureau, and Chengye Cao for helpful suggestions. This work is supported by the National Natural Science Foundation of China under grant No. 12233001, by  the National Key R$\&$D Program of China under grant No. 2024YFA1611602, by a Shanghai Natural Science Research Grant (24ZR1491200), by the ``111'' project of the Ministry of Education under grant No. B20019, and by the China Manned Space Project with No. CMS-CSST-2025-A09 and No. CMS-CSST-2025-A08. We thank the sponsorship from Yangyang Development Fund. Yang A. Li acknowledges the China Postdoctoral Science Foundation (No. 2023M742285) and the Postdoctoral Fellowship Program of CPSF (GZC20231611). This work made use of the Gravity Supercomputer at the Department of Astronomy, Shanghai Jiao Tong University.

\end{acknowledgements}

%

\bibliographystyle{aa} 
\bibliography{name} 







   
  



\begin{appendix}



\onecolumn

\section{Variation of Bulge Parameters with Different Barlens Sérsic Index} \label{apx}

Given that fixing the barlens Sérsic index during the 2D image decomposition is a simplified approach, we examine how fixing the Sérsic index to different values impacts the bulge parameters and explore the potential range of these parameters. 

\begin{figure*}[h]
\includegraphics[width=\textwidth]{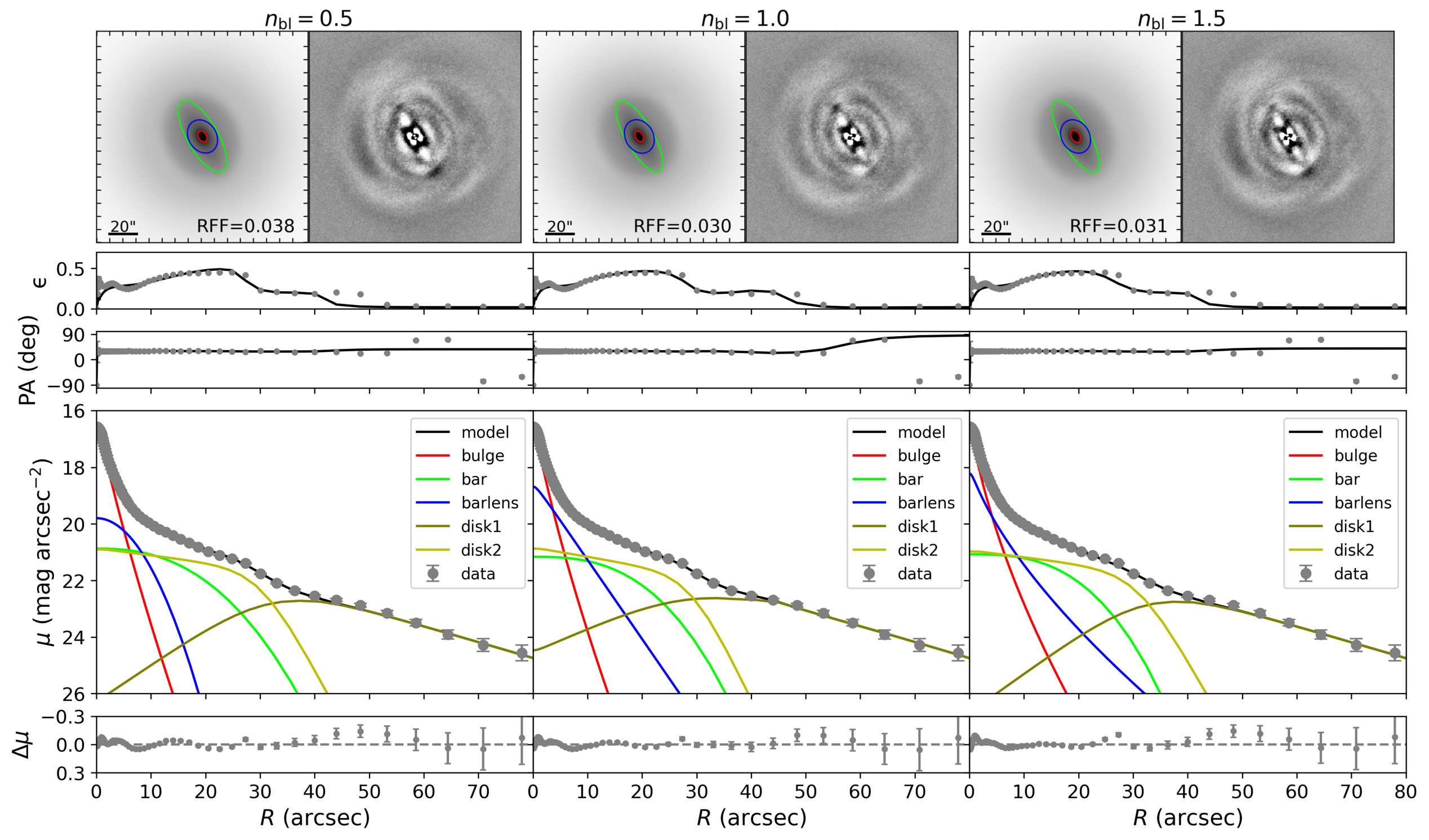}
    \caption{2D image decomposition results of fixing different $n_{\rm bl}$ in B+bar+bl+D1+D2 for Model 2. Other conventions are similar to Figure~\ref{fig:nbody_decomposition}. The value of $n_{\rm bl}$ fixed in the fitting is indicated on the model image in each column. }
    \label{fig:au6_nbl}
\end{figure*}

The results of Model 2 are shown in Figure~\ref{fig:au6_nbl} and listed in Table~\ref{tab:au6_differentnbl}. Increasing $n_{\rm bl}$ from 0.5 to 1.5 leads to an increase in the Sérsic index $n_{bulge}$ from 1.23 to 1.63. B/T remains largely unchanged ($\sim0.1$), while bl/T stays approximately 0.14 for $n_{\rm bl}=1.0$ and 1.5 but drops to $\sim0.09$ for $n_{\rm bl}=0.5$. The bar parameters are also not severely affected and remain a flat light profile, with the Sérsic index $\sim0.4$ and bar/T $\sim0.07$. Moreover, for B+bar+bl+D1+D2 (geo. fixed, $n_{\rm bl}=1.0$ and 1.5), RFF is $\sim0.030$, lower than the value in B+bar+D1+D2, indicating the reliability of the fitting results. However, for B+bar+bl+D1+D2 (geo. fixed, $n_{\rm bl}=0.5$), RFF is higher than the that in B+bar+D1+D2, suggesting that this is not a successful fitting configuration. This may indicate that the intrinsic value of $n_{\rm bl}$ for Model 2 lies between 1.0 and 1.5. 

\begin{figure*}[t]
\includegraphics[width=\textwidth]{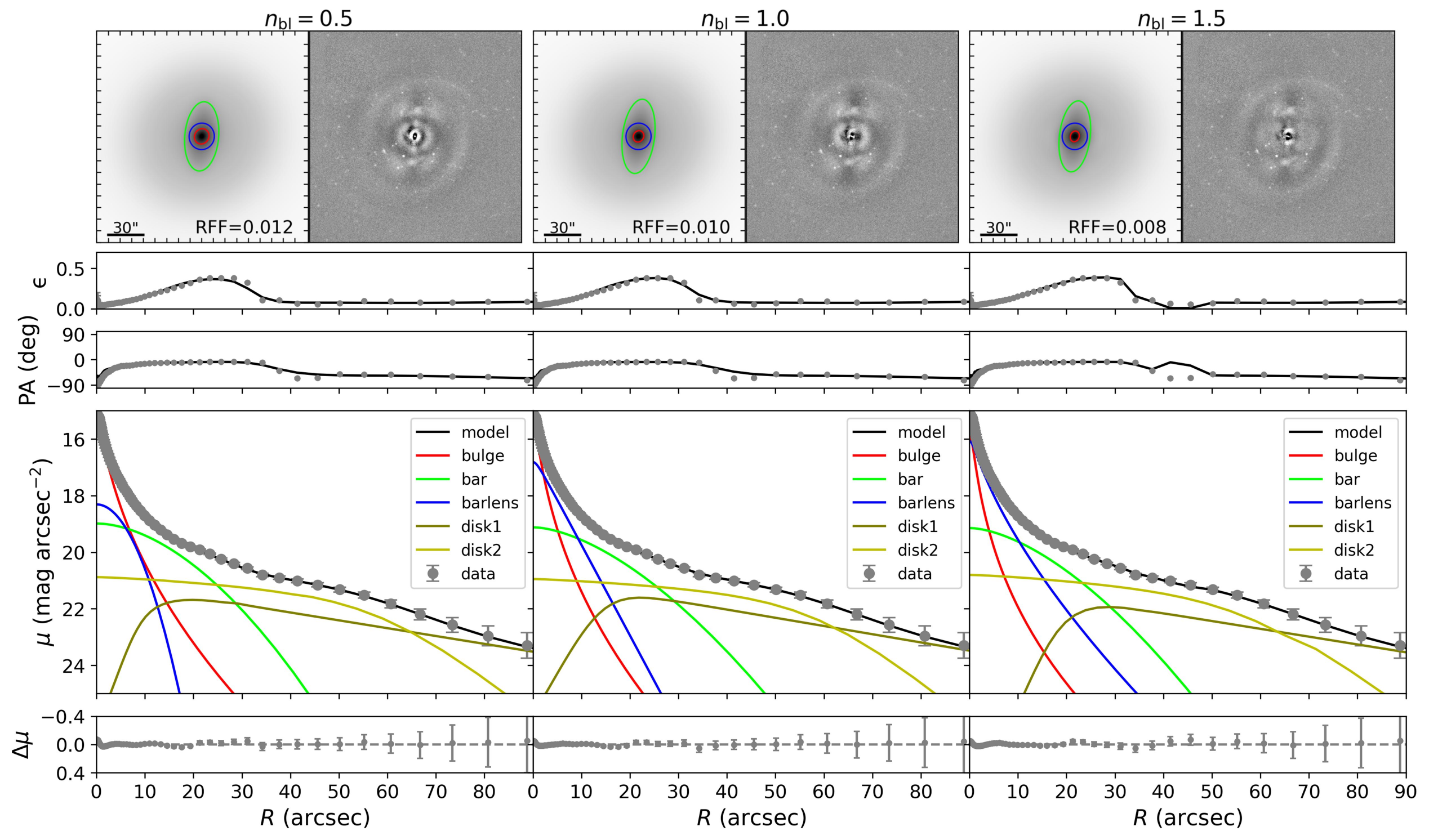}
    \caption{2D image decomposition results of fixing different $n_{\rm bl}$ in B+bar+bl+D1+D2 for NGC 1533. Other conventions are similar to Figure~\ref{fig:nbody_decomposition}. }
    \label{fig:NGC1533_nbl}
\end{figure*}

Figure~\ref{fig:NGC1533_nbl} shows the results of NGC 1533 and the corresponding parameters are listed in Table~\ref{tab:ngc1533_differentnbl}. Unlike in Model 2, the bulge parameters in NGC 1533 are significantly affected by the variation in $n_{\rm bl}$. B/T at $n_{\rm bl}=0.5$ ($\sim0.18$) is three times that at $n_{\rm bl}=1.5$ ($\sim0.06$). Compared to B+bar+D1+D2 (B/T$\sim0.33$), this represents a decrease of $40\%$ to $80\%$. The bulge Sérsic index ranges from 2.06 to 2.82, decreasing at $n_{\rm bl}=0.5$ and 1.0 and increasing at $n_{\rm bl}=1.5$ compared to B+bar+D1+D2. bl/T shows an opposite trend to B/T, increasing from 0.06 to 0.25 as $n_{\rm bl}$ increases. The bar component is minimally affected by the different barlens models, in contrast to the bulge and barlens components. Although RFF decreases compared to B+bar+D1+D2 in all three cases, $n_{\rm bl}=0.5$ results in the most unsatisfactory residual patterns in the central region of the residual map, suggesting that the intrinsic $n_{\rm bl}$ for NGC 1533 is likely between 1.0 and 1.5.

\begin{figure*}[ht]
\includegraphics[width=\textwidth]{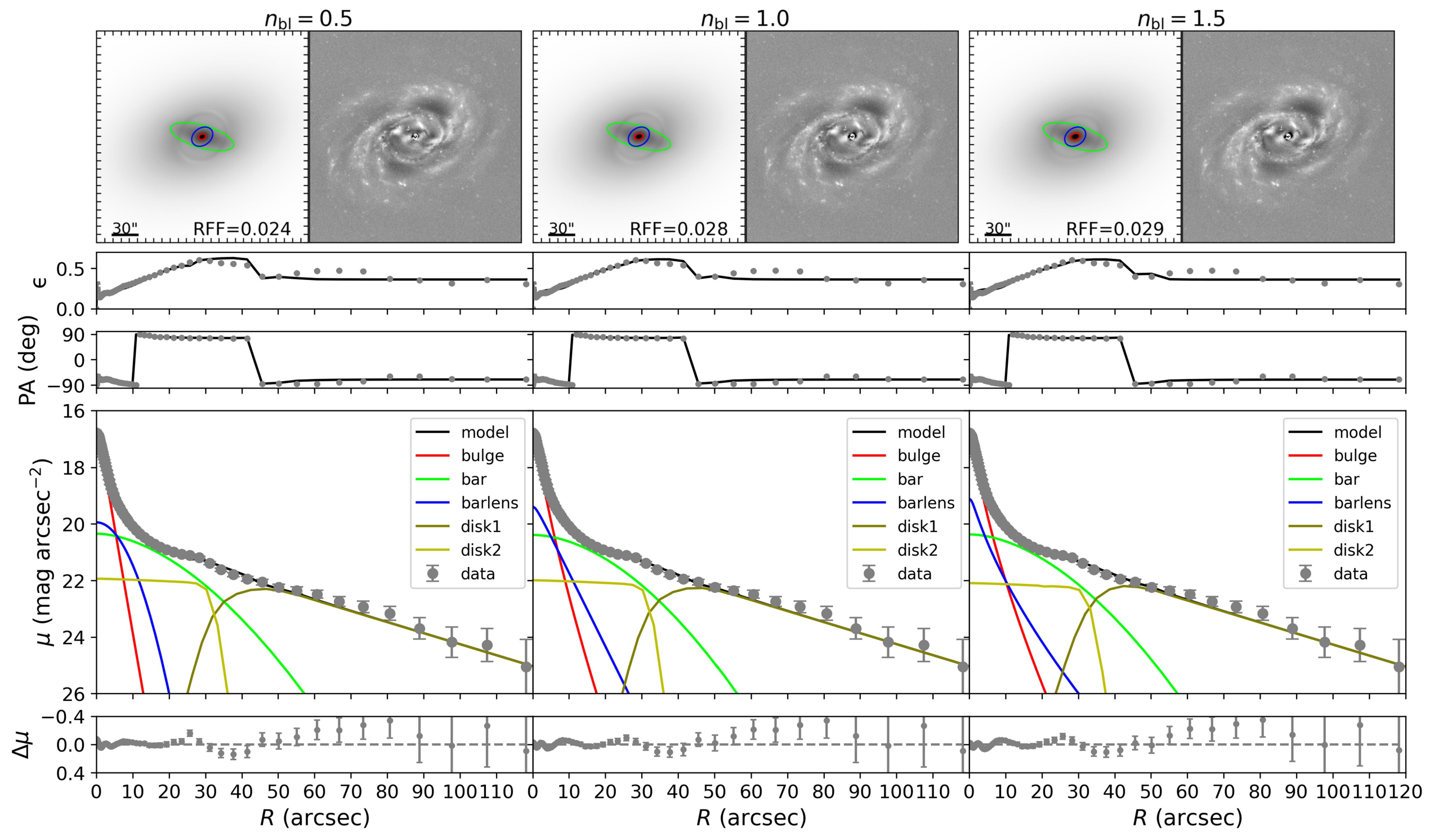}
    \caption{2D image decomposition results of fixing different $n_{\rm bl}$ in B+bar+bl+D1+D2 for NGC 7329. Other conventions are similar to Figure~\ref{fig:nbody_decomposition}. }
    \label{fig:NGC7329_nbl}
\end{figure*}

Figure~\ref{fig:NGC7329_nbl} shows the results of NGC 7329 and the corresponding parameters are listed in Table~\ref{tab:ngc7329_differentnbl}. Similar to Model 2, the bulge Sérsic index increases moderately from 1.04 to 1.52 as $n_{\rm bl}$ increase from 0.5 to 1.5 and B/T remains essentially unaffected ($\sim0.11$). Moreover, Additionally, variations in $n_{\rm bl}$ have minimal impact on the barlens parameters, with bl/T remaining around 0.07 and the bar parameters showing negligible change. However, for both $n_{\rm bl}=1.0$ and 1.5, RFF is similar to that in B+bar+D1+D2, suggesting that these fittings do not adequately capture the light distribution in the barlens region. This may indicate that the barlens in NGC 7329 has a relatively flat profile, which can be partly attributed to the inclination effect.

\begin{table*}[h]
	\caption{Best-fit parameters of fixing different $n_{\rm bl}$ in B+bar+bl+D1+D2 from 2D image decomposition of Model 2. The meaning of each column is shown in Table~\ref{tab:nbody_2D_params}. }
\label{tab:au6_differentnbl}
\resizebox{517pt}{!}{
	\begin{tabular}{cccccccc}
		\hline
		Component & Configuration & $n$ & $r_{\rm e}$ & $\mu_{\rm e}$ & b/a & PA & $f/f_{\rm tot}$ \\
             &  &  & (arcsec) & (mag arcsec$^{-2}$) &  & ($^\circ$) &  \\
             (1) & (2) & (3) & (4) & (5) & (6) & (7) & (8) \\ 
		\hline
		\multirow{3}{*}{bulge} & B+bar+bl+D1+D2 (geo. fixed, $n_{\rm bl}$=0.5) & $1.23 \pm 0.05$ & $2.24 \pm 0.06$ & $18.08 \pm 0.06$ & $0.68 \pm 0.01$ & $29.62\pm 0.08$ & $0.12\pm 0.01$ \\
        & B+bar+bl+D1+D2 (geo. fixed, $n_{\rm bl}$=1.0) & $1.34 \pm 0.04$ & $2.10 \pm 0.06$ & $18.19 \pm 0.05$ & $0.65 \pm 0.01$ & $29.06 \pm 0.08$ & $0.10 \pm 0.01$ \\
        & B+bar+bl+D1+D2 (geo. fixed, $n_{\rm bl}$=1.5) & $1.63 \pm 0.05$ & $2.54 \pm 0.07$ & $18.61\pm 0.09$ & $0.64 \pm 0.01$ & $29.54 \pm 0.06$ & $0.10 \pm 0.01$ \\
        \hline
		\multirow{3}{*}{bar} & B+bar+bl+D1+D2 (geo. fixed, $n_{\rm bl}$=0.5) & $0.41\pm 0.03$ & $14.95\pm 0.33$ & $21.43\pm 0.04$ & $0.38\pm 0.01$ & $29.61\pm 0.05$ & $0.09\pm 0.01$ \\
        & B+bar+bl+D1+D2 (geo. fixed, $n_{\rm bl}$=1.0) & $0.36\pm 0.03$ & $15.17 \pm 0.35$ & $21.63 \pm 0.03$ & $0.31 \pm 0.01$ & $29.78 \pm 0.06$ & $0.06 \pm 0.01$ \\
        & B+bar+bl+D1+D2 (geo. fixed, $n_{\rm bl}$=1.5) & $0.34\pm 0.04$ & $15.10\pm 0.36$ & $21.51\pm 0.04$ & $0.34\pm 0.01$ & $29.93\pm 0.06$ & $0.07\pm 0.01$ \\
        \hline
        \multirow{3}{*}{barlens} & B+bar+bl+D1+D2 (geo. fixed, $n_{\rm bl}$=0.5) & $0.50\pm 0.10$(fixed) & $6.50\pm 0.63$(fixed) & $20.53 \pm 0.04$ & $0.79\pm 0.03$(fixed) & $33.23 \pm 0.14$ & $0.09 \pm 0.01$ \\
        & B+bar+bl+D1+D2 (geo. fixed, $n_{\rm bl}$=1.0) & $1.00\pm 0.10$(fixed) & $6.50\pm 0.63$(fixed) & $20.29 \pm 0.03$ & $0.79\pm 0.03$(fixed) & $32.94 \pm 0.11$ & $0.14 \pm 0.02$ \\
        & B+bar+bl+D1+D2 (geo. fixed, $n_{\rm bl}$=1.5) & $1.50\pm 0.10$(fixed) & $6.50\pm 0.63$(fixed) & $20.48 \pm 0.03$ & $0.79\pm 0.03$(fixed) & $31.76 \pm 0.09$ & $0.14 \pm 0.02$ \\
	\hline
	\end{tabular}
}
\end{table*}

\begin{table*}[t]
	\caption{Best-fit parameters of fixing different $n_{\rm bl}$ in B+bar+bl+D1+D2 from 2D image decomposition of NGC 1533. The meaning of each column is shown in Table~\ref{tab:nbody_2D_params}. The value of $n_{\rm bl}$ fixed in the fitting is indicated on the model image in each column. }
\label{tab:ngc1533_differentnbl}
\resizebox{517pt}{!}{
	\begin{tabular}{cccccccc}
		\hline
		Component & Configuration & $n$ & $r_{\rm e}$ & $\mu_{\rm e}$ & b/a & PA & $f/f_{\rm tot}$ \\
             &  &  & (arcsec) & (mag arcsec$^{-2}$) &  & ($^\circ$) &  \\
             (1) & (2) & (3) & (4) & (5) & (6) & (7) & (8) \\ 
		\hline
		\multirow{3}{*}{bulge} & B+bar+bl+D1+D2 (geo. fixed, $n_{\rm bl}$=0.5) & $2.06 \pm 0.14$ & $3.37 \pm 0.14$ & $17.58 \pm 0.03$ & $0.90 \pm 0.01$ & $-29.09\pm 9.58$ & $0.18\pm 0.02$\\
        & B+bar+bl+D1+D2 (geo. fixed, $n_{\rm bl}$=1.0) & $2.32\pm 0.12$ & $2.53\pm 0.19$ & $17.63\pm 0.02$ & $0.87\pm 0.03$ & $-27.71\pm13.01$ & $0.10\pm0.02$ \\
        & B+bar+bl+D1+D2 (geo. fixed, $n_{\rm bl}$=1.5) & $2.82 \pm 0.15$ & $2.34 \pm 0.15$ & $18.08\pm 0.09$ & $0.80 \pm 0.01$ & $-21.67 \pm 11.36$ & $0.06 \pm 0.01$ \\
        \hline
		\multirow{3}{*}{bar} & B+bar+bl+D1+D2 (geo. fixed, $n_{\rm bl}$=0.5) & $0.56\pm 0.03$ & $14.89\pm 0.17$ & $19.85\pm 0.08$ & $0.49\pm 0.01$ & $-5.43\pm 0.10$ & $0.13\pm 0.01$ \\
        & B+bar+bl+D1+D2 (geo. fixed, $n_{\rm bl}$=1.0) & $0.61\pm 0.05$ & $16.01\pm 0.26$ & $20.09\pm 0.10$ & $0.43\pm 0.01$ & $-7.07\pm0.22$ & $0.11\pm0.01$ \\
        & B+bar+bl+D1+D2 (geo. fixed, $n_{\rm bl}$=1.5) & $0.61\pm 0.04$ & $15.32\pm 0.19$ & $20.11\pm 0.09$ & $0.43\pm 0.01$ & $-7.02\pm 0.22$ & $0.10\pm 0.01$ \\
        \hline
        \multirow{3}{*}{barlens} & B+bar+bl+D1+D2 (geo. fixed, $n_{\rm bl}$=0.5) & $0.50\pm 0.10$(fixed) & $5.68\pm 0.84$(fixed) & $19.03 \pm 0.25$ & $0.95\pm 0.04$(fixed) & $-88.34 \pm 2.66$ & $0.08 \pm 0.01$ \\
        & B+bar+bl+D1+D2 (geo. fixed, $n_{\rm bl}$=1.0) & $1.00\pm0.10$ (fixed) & $5.68\pm 0.84$ (fixed) & $18.37\pm0.41$ & $0.95\pm 0.04$ (fixed)& $-76.67\pm2.34$ & $0.18\pm 0.02$ \\
        & B+bar+bl+D1+D2 (geo. fixed, $n_{\rm bl}$=1.5) & $1.50\pm 0.10$(fixed) & $5.68\pm 0.84$(fixed) & $18.24 \pm 0.43$ & $0.95\pm 0.04$ (fixed) & $-69.11 \pm 3.95$ & $0.25 \pm 0.02$ \\
	\hline
	\end{tabular}
}
\end{table*}

\begin{table*}[t!]
	\caption{Best-fit parameters of fixing different $n_{\rm bl}$ in B+bar+bl+D1+D2 from 2D image decomposition of NGC 7329. The meaning of each column is shown in Table~\ref{tab:nbody_2D_params}. The value of $n_{\rm bl}$ fixed in the fitting is indicated on the model image in each column. }
\label{tab:ngc7329_differentnbl}
\resizebox{517pt}{!}{
	\begin{tabular}{cccccccc}
		\hline
		Component & Configuration & $n$ & $r_{\rm e}$ & $\mu_{\rm e}$ & b/a & PA & $f/f_{\rm tot}$ \\
             &  &  & (arcsec) & (mag arcsec$^{-2}$) &  & ($^\circ$) &  \\
             (1) & (2) & (3) & (4) & (5) & (6) & (7) & (8) \\ 
		\hline
		\multirow{3}{*}{bulge} & B+bar+bl+D1+D2 (geo. fixed, $n_{\rm bl}$=0.5) & $1.04\pm 0.04$ & $2.31\pm 0.07$ & $18.09\pm 0.04$ & $0.74\pm0.01$ & $ -74.77\pm 2.36$ & $0.11\pm 0.02$ \\
        & B+bar+bl+D1+D2 (geo. fixed, $n_{\rm bl}$=1.0) & $1.38\pm 0.05$ & $2.72\pm 0.33$ & $18.44\pm 0.15$ & $0.70\pm0.01$ & $ -77.34\pm 2.71$ & $0.11\pm 0.02$ \\
        & B+bar+bl+D1+D2 (geo. fixed, $n_{\rm bl}$=1.5) & $1.52\pm 0.07$ & $3.13\pm 0.39$ & $18.65\pm 0.13$ & $0.67\pm0.01$ & $ -76.82\pm 1.86$ & $0.12\pm 0.02$ \\
        \hline
		\multirow{3}{*}{bar} & B+bar+bl+D1+D2 (geo. fixed, $n_{\rm bl}$=0.5) & $0.57\pm 0.17$ & $19.98\pm 1.02$ & $21.23\pm 0.07$ & $0.35\pm 0.01$ & $74.31\pm0.51$ & $0.16\pm0.05$ \\
        & B+bar+bl+D1+D2 (geo. fixed, $n_{\rm bl}$=1.0) & $0.55\pm 0.13$ & $19.99\pm 0.34$ & $21.21\pm 0.07$ & $0.34\pm 0.01$ & $74.37\pm0.25$ & $0.15\pm0.04$ \\
        & B+bar+bl+D1+D2 (geo. fixed, $n_{\rm bl}$=1.5) & $0.56\pm 0.16$ & $20.21\pm 0.35$ & $21.23\pm 0.07$ & $0.34\pm 0.01$ & $74.38\pm0.31$ & $0.16\pm0.04$ \\
        \hline
        \multirow{3}{*}{barlens} & B+bar+bl+D1+D2 (geo. fixed, $n_{\rm bl}$=0.5) & $0.50\pm0.10$ (fixed) & $6.99\pm 0.78$ (fixed) & $20.67\pm 0.06$ & $0.75\pm 0.03$ (fixed) & $-58.11\pm 1.18$ & $0.07\pm 0.02$ \\
        & B+bar+bl+D1+D2 (geo. fixed, $n_{\rm bl}$=1.0) & $1.00\pm0.10$ (fixed) & $6.99\pm 0.78$ (fixed) & $20.95\pm 0.37$ & $0.75\pm 0.03$ (fixed) & $-45.55\pm 1.91$ & $0.07\pm 0.02$ \\
        & B+bar+bl+D1+D2 (geo. fixed, $n_{\rm bl}$=1.5) & $1.50\pm0.10$ (fixed) & $6.99\pm 0.78$ (fixed) & $21.23\pm 0.50$ & $0.75\pm 0.03$ (fixed) & $-29.43\pm 1.35$ & $0.06\pm 0.01$ \\
	\hline
	\end{tabular}
}
\end{table*}

\section{Decomposition with a Gaussian Ring} \label{apx2}

Since modeling the ring structure with two broken exponential disks increases the number of free parameters and may introduce degeneracies between the inner disk and the barlens, we tested an alternative approach using a Gaussian ring combined with an exponential disk to represent the ring structure in Model 2. Specifically, we adopted the configurations B+bar+R+D and B+bar+bl+R+D, where R denotes the Gaussian ring. As GALFIT does not support a Gaussian ring component, we performed the fitting using IMFIT \citep{Erwin2015}, which provides this functionality. To ensure consistency, we employed the same optimization algorithm used in GALFIT, namely the Levenberg–Marquardt algorithm. As in previous fits, we fixed the barlens Sérsic index to 0.5, 1.0, and 1.5, and selected the result with the lowest RFF as the final fit. 

\begin{figure*}[h]
\includegraphics[width=\textwidth]{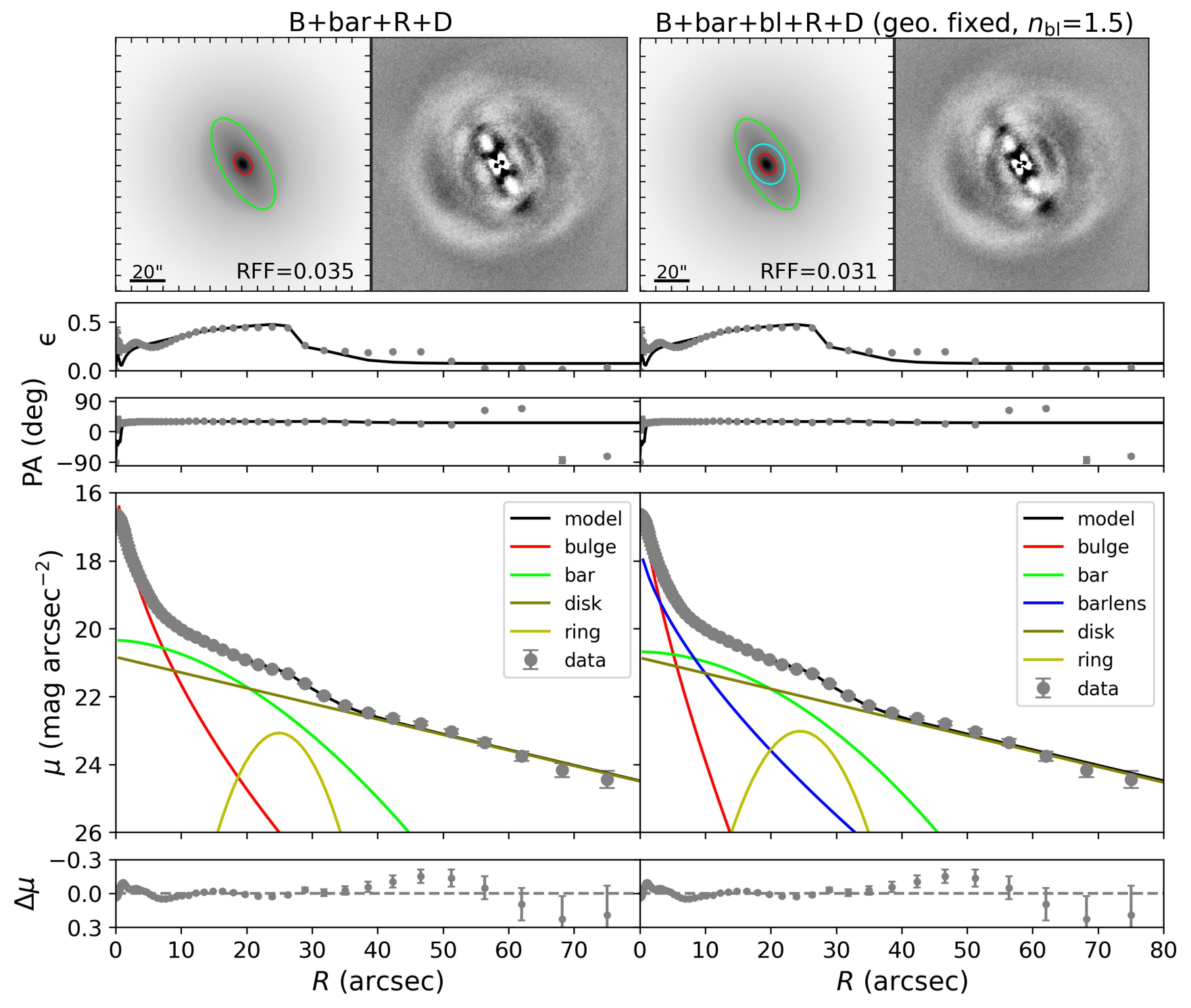}
    \caption{2D image decomposition results of using the Gaussian ring profile to model the ring structure in B+bar+R+D and B+bar+bl+R+D for Model 2. Other conventions are similar to Figure~\ref{fig:nbody_decomposition}. }
    \label{fig:gaussianring}
\end{figure*}

The results are shown in Figure~\ref{fig:gaussianring} and listed in Table~\ref{tab:model2_gaussianring}.
Compared to the configurations using broken exponential disks, those employing a Gaussian ring yield similar RFF values and residual maps. Overall, the fitting results are consistent between the two modeling approaches, particularly for the bulge parameters and after the inclusion of the barlens component. When the Gaussian ring is adopted, the bar shows a slightly higher Sérsic index and effective radius. Unlike the case with broken exponential disks, the best-fit Sérsic index for the barlens is 1.5; however, bl/T remains nearly unchanged. Considering that the choice of ring model has no significant impact on the results, we adopt the broken exponential disk in the final analysis to facilitate direct comparison with \citet{Gao2017}.

\begin{table*}[t]
	\caption{Best-fit parameters of modeling a Gaussian ring in B+bar+R+D and B+bar+bl+R+D from 2D image decomposition of Model 2. The meaning of each column is shown in Table~\ref{tab:nbody_2D_params}. }
\label{tab:model2_gaussianring}
\resizebox{517pt}{!}{
	\begin{tabular}{cccccccc}
		\hline
		Component & Configuration & $n$ & $r_{\rm e}$ & $\mu_{\rm e}$ & b/a & PA & $f/f_{\rm tot}$ \\
             &  &  & (arcsec) & (mag arcsec$^{-2}$) &  & ($^\circ$) &  \\
             (1) & (2) & (3) & (4) & (5) & (6) & (7) & (8) \\ 
		\hline
		\multirow{2}{*}{bulge} & B+bar+R+D & $1.77 \pm 0.18$ & $3.39 \pm 0.20$ & $18.71 \pm 0.15$ & $0.79 \pm 0.01$ & $30.49\pm 0.42$ & $0.18\pm 0.02$\\
        & B+bar+bl+R+D (geo. fixed, $n_{\rm bl}$=1.5) & $1.39 \pm 0.06$ & $2.10 \pm 0.05$ & $18.36\pm 0.04$ & $0.67 \pm 0.01$ & $28.88 \pm 0.06$ & $0.08 \pm 0.01$ \\
        \hline
		\multirow{2}{*}{bar} & B+bar+R+D & $0.57\pm 0.03$ & $15.70\pm 0.31$ & $21.24\pm 0.06$ & $0.43\pm 0.01$ & $29.67\pm 0.11$ & $0.14\pm 0.01$ \\
        & B+bar+bl+R+D (geo. fixed, $n_{\rm bl}$=1.5) & $0.52\pm 0.03$ & $16.92\pm 0.30$ & $21.48\pm 0.04$ & $0.33\pm 0.01$ & $29.36\pm 0.06$ & $0.09\pm 0.01$ \\
        \hline
        \multirow{1}{*}{barlens} & B+bar+bl+R+D (geo. fixed, $n_{\rm bl}$=1.5) & $1.50\pm 0.10$(fixed) & $6.50\pm 0.63$(fixed) & $20.34 \pm 0.07$ & $0.79\pm 0.03$ (fixed) & $33.00 \pm 0.25$ & $0.15 \pm 0.02$ \\
	\hline
	\end{tabular}
}
\end{table*}

\section{Best-fit Disk Parameters from the 2D Image Decomposition} \label{apx3}

Table~\ref{tab:disk} summarizes the best-fit parameters of the disk component for each model configuration applied to the mock images and the two CGS galaxies analyzed in this study. For configurations involving two broken exponential disks, the parameters of the inner and outer disks are listed separately.

For Model 1, when comparing the bar+D and bl+bar+D configurations, the disk scale length slightly decreases after including the barlens component. However, in the case of bar+D1+D2 versus bl+bar+D1+D2, the scale length of the inner disk increases. Additionally, the inner disk in the broken exponential profile exhibits a flatter slope compared to that of the single exponential disk. The central surface brightness of the inner disk and the parameters of the outer disk remain roughly unchanged. For Model 2, the changes in disk parameters after including the barlens component are similar to those observed in Model 1, except that the scale length of the inner disk is slightly reduced.

The inclusion of the barlens component raises the inner disk scale length more significantly for NGC 1533, whereas all the disk parameters remain roughly unchanged for NGC 7329.

\begin{table*}[t]
	\caption{Best-fit parameters of the disk component of different galaxies and different configurations from 2D image decomposition. The columns from left to right show the data image, the decomposition configuration, the scale length, the central surface brightness, the axial ratio, the position angle and the disk-to-total flux ratio, respectively. }
\label{tab:disk}
\resizebox{517pt}{!}{
	\begin{tabular}{cccccccc}
\hline
    Data & Configuration & inner/outer disk & $r_{\rm h}$ & $\mu_{0}$ & b/a & PA & D/T \\
          &           &                    & (arcsec)    & (mag arcsec$^{-2}$) & & ($^\circ$) &  \\
    (1)   & (2)       & (3)           & (4) & (5)         & (6)         & (7) & (8)         \\
    \hline
    \multirow[c]{8}{*}{Model 1} 
        & \multirow[l]{1}{*}{bar+D} & / & $24.62\pm6.61$ & $20.44\pm0.45$ & $0.93\pm0.11$ & $-5.30\pm3.79$ & $0.55\pm0.05$ \\
    \cline{2-8}
        & \multirow[l]{2}{*}{bar+D1+D2}  & inner & $28.31\pm3.09$  & $21.05\pm0.22$ & $0.90\pm0.01$ & $3.14\pm4.23$ & $0.19\pm0.03$ \\
        &                         & outer & $13.86\pm6.56$ & $17.99\pm1.23$ & $0.95\pm0.01$ & $89.38\pm4.14$  & $0.33\pm0.06$ \\
    \cline{2-8}
        & \multirow[l]{2}{*}{bl+bar+D1+D2 (free)}  & inner & $37.25\pm4.21$  & $21.16\pm0.31$ & $0.90\pm0.01$ & $-2.37\pm3.86$ & $0.22\pm0.02$ \\
        &                         & outer & $12.95\pm6.98$ & $17.57\pm1.87$ & $0.94\pm0.01$ & $89.79\pm3.87$  & $0.30\pm0.06$ \\
    \cline{2-8}
        & \multirow[l]{2}{*}{bl+bar+D1+D2 (geo. fixed)}  & inner & $38.75\pm3.09$  & $21.05\pm0.29$ & $0.72\pm0.01$ & $-1.97\pm1.39$ & $0.22\pm0.02$ \\
        &                         & outer & $12.92\pm7.02$ & $17.55\pm1.87$ & $0.94\pm0.01$ & $91.51\pm3.86$  & $0.31\pm0.06$ \\
    \cline{2-8}
        & \multirow[l]{1}{*}{bl+bar+D} & / & $23.98\pm4.08$ & $20.48\pm0.19$ & $0.98\pm0.04$ & $-37.35\pm33.49$ & $0.53\pm0.06$ \\
    \hline
    \multirow[c]{10}{*}{Model 2}
        & \multirow[l]{1}{*}{B+D} & / & $22.46\pm6.64$ & $20.92\pm0.83$ & $0.96\pm0.07$ & $-60.87\pm26.94$ & $0.55\pm0.14$ \\
    \cline{2-8}
        & \multirow[l]{1}{*}{B+bar+D} & / & $21.59\pm6.10$ & $20.63\pm0.62$ & $0.96\pm0.06$ & $30.72\pm3.53$ & $0.67\pm0.04$ \\
    \cline{2-8}
        & \multirow[l]{2}{*}{B+bar+D1+D2}  & inner & $37.59\pm5.96$  & $20.76\pm0.62$ & $0.60\pm0.04$ & $28.88\pm0.06$ & $0.31\pm0.10$ \\
        &                         & outer & $18.95\pm10.46$ & $20.17\pm0.71$ & $0.98\pm0.01$ & $81.79\pm21.50$  & $0.39\pm0.09$ \\
    \cline{2-8}
        & \multirow[l]{2}{*}{B+bar+bl+D1+D2 (free)}  & inner & $39.08\pm4.02$  & $20.98\pm0.07$ & $0.84\pm0.09$ & $30.02\pm0.15$ & $0.29\pm0.01$ \\
        &                         & outer & $19.26\pm7.43$ & $20.24\pm2.70$ & $0.98\pm0.04$ & $32.83\pm7.55$  & $0.40\pm0.03$ \\
    \cline{2-8}
        & \multirow[l]{2}{*}{B+bar+bl+D1+D2 (geo. fixed)}  & inner & $38.08\pm1.04$  & $20.95\pm0.15$ & $0.71\pm0.01$ & $28.67\pm3.76$ & $0.28\pm0.01$ \\
        &                         & outer & $19.65\pm7.25$ & $20.32\pm0.55$ & $0.98\pm0.02$ & $36.45\pm10.04$  & $0.39\pm0.03$ \\
    \cline{2-8}
        & \multirow[l]{2}{*}{B+bar+bl+D1+D2 (geo. fixed, $n_{\rm bl}$=1.0)}  & inner & $32.57\pm8.40$  & $20.83\pm1.37$ & $0.68\pm0.01$ & $30.11\pm5.46$ & $0.27\pm0.05$ \\
        &                         & outer & $18.73\pm7.40$ & $20.10\pm1.02$ & $0.98\pm0.02$ & $86.23\pm57.46$  & $0.43\pm0.01$ \\
    \hline
    \multirow[c]{6}{*}{NGC 1533}
        & \multirow[l]{1}{*}{B+D} & / & $25.31\pm7.18$ & $19.37\pm0.38$ & $0.93\pm0.13$ & $-47.76\pm7.24$ & $0.69\pm0.02$ \\
    \cline{2-8}
        & \multirow[l]{1}{*}{B+bar+D} & / & $25.35\pm4.08$ & $19.37\pm0.23$ & $0.93\pm0.01$ & $-52.58\pm9.94$ & $0.70\pm0.03$ \\
    \cline{2-8}
        & \multirow[l]{2}{*}{B+bar+D1+D2}  & inner & $28.84\pm8.54$  & $20.32\pm0.08$ & $0.91\pm0.07$ & $-44.87\pm15.51$ & $0.26\pm0.01$ \\
        &           & outer & $32.86\pm9.62$ & $20.39\pm0.16$ & $0.91\pm0.03$ & $-69.36\pm6.14$  & $0.32\pm0.10$ \\
    \cline{2-8}
        & \multirow[l]{2}{*}{B+bar+bl+D1+D2 (geo. fixed, $n_{\rm bl}$=1.5)}  & inner & $54.95\pm15.13$  & $20.68\pm0.14$ & $0.91\pm0.01$ & $-47.68\pm3.92$ & $0.36\pm0.07$ \\
        &             & outer & $39.21\pm12.41$ & $21.05\pm0.49$ & $0.89\pm0.02$ & $-74.03\pm17.89$  & $0.24\pm0.11$ \\
    \hline   
    \multirow[c]{6}{*}{NGC 7329}
        & \multirow[l]{1}{*}{B+D} & / & $29.64 \pm 9.52$ & $20.59 \pm 0.37$ & $0.64\pm 0.01$ & $-74.98\pm0.83$ & $0.78\pm0.02$ \\
    \cline{2-8}
        & \multirow[l]{1}{*}{B+bar+D} & / & $29.64\pm7.61$ & $20.67\pm0.47$ & $0.65\pm0.04$ & $-71.05\pm2.07$ & $0.73\pm0.02$ \\
    \cline{2-8}
        & \multirow[l]{2}{*}{B+bar+D1+D2}  & inner & $98.78\pm15.54$  & $21.86\pm0.03$ & $0.75\pm0.01$ & $-71.00\pm1.00$ & $0.17\pm0.01$ \\
        &                         & outer & $26.70\pm8.46$ & $20.28\pm0.24$ & $0.64\pm0.01$ & $-70.50\pm3.40$  & $0.55\pm0.08$ \\
    \cline{2-8}
        & \multirow[l]{2}{*}{B+bar+bl+D1+D2 (geo. fixed, $n_{\rm bl}$=0.5)}  & inner & $92.29\pm15.50$ & $21.74\pm0.02$ & $0.65\pm0.01$ & $-71.00\pm1.00$ & $0.18\pm0.01$ \\
        &                         & outer & $26.38\pm8.31$ & $20.23\pm0.25$ & $0.64\pm0.01$ & $-70.52\pm3.15$  & $0.54\pm0.08$ \\
    \hline   
\end{tabular}
}
\end{table*}

\end{appendix}

\end{document}